\documentclass[pra,
,tightenlines
,showkeys
,showpacs
]
{revtex4-2}

\usepackage{amsmath,amssymb}
\usepackage{bm}
\usepackage[dvipdfmx]{graphicx}
\usepackage{ascmac} 
\usepackage{fancybox}
\usepackage{float}
\usepackage{enumerate}
\usepackage{color}
\usepackage{amsthm}
\usepackage{mathrsfs}
\usepackage{comment}
\usepackage{ulem}
\allowdisplaybreaks[1]
\usepackage{comment}
\newtheorem*{th.}{Theorem}

\newcommand{\vsigma}{\vec{\sigma}}

\newcommand{\vt}{{\cal T}}

\begin{document}
	\title{
		General off-resonance error robust symmetric composite pulses with three elementary operations
	}
	\date{\today}
	\author{Shingo Kukita$^{1)}$}
	\email{toranojoh@phys.kindai.ac.jp}
	\author{Haruki Kiya$^{1)}$}
	\email{kiya.haruki@kindai.ac.jp}
	\author{Yasushi Kondo$^{1)}$}
	\email{ykondo@kindai.ac.jp}
	\affiliation{$^{1)}$Department of Physics, 
		Kindai University, Higashi-Osaka 577-8502, Japan}
	\begin{abstract}
		
	Accurate quantum control is a key technology for realizing quantum information processing, such as quantum communication and quantum computation.
	In reality, a quantum state under control suffers from undesirable effects caused by systematic errors.
	A composite pulse (CP) is used to eliminate the effects of systematic errors during control.
	One qubit control, which is the most fundamental in quantum control, is typically affected by two errors: pulse length error (PLE) and off-resonance error (ORE).
	In this study, we focus on ORE-robust CPs and systematically construct ORE-robust symmetric CPs with three elementary operations.
	We find an infinitely large number of ORE-robust CPs and evaluate their performance according to gate infidelity and operation time, both of which are important for the realization of accurate quantum control.
	
	\end{abstract}
	
	\maketitle
	
	\section{introduction}
	
	The performance of quantum technologies such as quantum computing \cite{Nielsen2000,bennett2000quantum,nakahara2008quantum}, quantum communication \cite{ekert1991quantum,gisin2007quantum,chen2021integrated}, and quantum metrology \cite{helstrom1976quantum,caves1981quantum,holevo2011probabilistic} is highly dependent on the accuracy of quantum control in each process.
	Numerous attempts have been made to improve the accuracy of quantum control \cite{RevModPhys.76.1037,PhysRevLett.99.036403,bason2012high,lidar2013quantum,chang2014band,spiteri2018quantum,levy2018noise}.
	Particularly, one-qubit control, which is the most fundamental part of quantum computation, has received considerable attention \cite{PhysRevA.73.022332,PhysRevLett.111.050404,PhysRevA.94.032323,yang2019achieving}.
	One-qubit control is also interesting from a geometrical perspective because this process is related to rotations on a unit sphere, which is called the Bloch sphere.
	
	Realistic quantum operations suffer from undesirable effects of systematic errors caused by the miscalibration of experimental apparatuses, and such errors deteriorate the performance of quantum control.
	In one-qubit control, one mainly confronts two typical systematic errors: pulse length error (PLE) and off-resonance error (ORE).
	PLE is an error due to the deviation in the external control field whereas ORE is often caused by the miscalibration of the resonance frequency of the qubit to be controlled. 
	These two errors in one-qubit control have been intensively investigated in the context of nuclear magnetic resonance (NMR), which can be used to demonstrate a toy quantum computer \cite{gershenfeld1997bulk,cory1997ensemble,vandersypen2001experimental,kondo2009liquid,JONES201191}.
	Even after we calibrate the parameters, slow and small fluctuations of the parameters can unavoidably cause these errors.	
	We then require treatments for the systematic errors without relying on calibration.
		
	A composite pulse (CP) is a method used to compensate for such systematic errors and has been investigated, particularly in the field of NMR \cite{counsell1985analytical,levitt1986composite,claridge2016high}.
	This method replaces a single operation (or pulse in the context of NMR) with a sequence of several operations such that the systematic errors in each operation cancel each other.
	Thus, the CPs are less-sensitive to these errors.
	For one-qubit operations, several CPs that are robust against PLE have been found, such as SK1 \cite{brown2004arbitrarily}, BB1 \cite{wimperis1994broadband}, and SCROFULOUS \cite{cummins2003tackling}.
	Similarly, there are several construction methods for ORE-robust CPs \cite{PhysRevA.83.053420,PhysRevLett.106.233001,jones2013designing,PhysRevLett.113.043001,PhysRevA.100.032333}.
	Particularly, when we implement a specific angle rotation in the Bloch sphere representation, such as $\pi$- and $\pi/2$- rotation, these methods work well and can provide simple and explicit formulae for determining ORE-robust operations.
	The CORPSE family is one of the simplest and most tractable CPs for implementing arbitrary $\theta$-rotations \cite{cummins2000use}; it does not require frequency modulation, and all of its parameters are explicitly determined as a simple function of the parameters of the target operation.
	This family often appears when we require ORE-robust arbitrary $\theta$-rotations \cite{mottonen2006high,said2009robust,timoney2008error,bando2012concatenated}.
	
	In this paper, we explain a systematic construction of a wide class of ORE-robust CPs that implement arbitrary $\theta$-rotations.
	The constructed class is time symmetric, consists of three elementary operations, and includes the CORPSE family.
	This construction provides the parameters in a CP as an explicit function of the parameters in the target operation as the CORPSE family does.
	Surprisingly, even in this restricted class of CPs, there exist an infinitely large number of CPs for arbitrary $\theta$-rotations:
	we have one continuous free parameter to choose a CP that implements a target operation. 
	We then evaluate the performance of the CPs in this class in terms of gate infidelity and the time required for operation.
	Small gate infidelity is necessary for accurate quantum control, whereas a short operation time is generally required, for example, to avoid the effect of environmental noise.
	As a result, we validate the excellent performance of the CORPSE family in this wide class of ORE-robust CPs.	
	
	The remainder of this paper is organized as follows. In Sec.~\ref{sec:II}, we briefly review the concept of CPs while focusing on ORE-robust ones.
	Sec.~\ref{sec:III} explains the systematic construction of ORE-robust CPs with three elementary operations, and show that there are a large number of ORE-robust CPs based on this construction.
	Then, we evaluate the CPs that we found in terms of gate fidelity and operation time in Sec.~\ref{sec:IIII}.
	Sec.~\ref{sec:IIIII} concludes the paper.

	\section{review of composite pulses}
	\label{sec:II}
	
	A one-qubit operation is described by a unitary matrix:
	\begin{equation}
		U(\theta,\vec{n}):=\exp \bigl( - i \theta \vec{n}\cdot\vsigma/2\bigr)=\cos(\theta/2){\mathbb I}- i \sin(\theta/2)\vec{n}\cdot \vsigma,~~\theta>0,
		\label{eq:elementary}
	\end{equation}
	where $\vec{n}$ is a three-dimensional unit vector and $\vsigma$ is the vector consisting of Pauli matrices, $\vsigma:=(\sigma_{x},\sigma_{y},\sigma_{z})$.
	Note that the positivity of $\theta$ results from the one-way time flow.
	In the Bloch sphere representation, a one-qubit state is represented by a point on the unit sphere; accordingly, a one-qubit operation is regarded as a rotation.
	In this representation, $\theta$ and $\vec{n}$ correspond to the angle and axis of rotation, respectively.
	Hereinafter, all operations that we consider are assumed to be constructed as a sequence of operations in the form of Eq. (\ref{eq:elementary}).
	Hence, we call this type of operation an ``elementary'' operation.
	Typically, for each elementary operation of a CP, $\vec{n}$ is considered to be in the $xy$ plane and parameterized by one parameter $\phi$ as $\vec{n}=(\cos\phi, \sin\phi,0)$.
	Hereinafter, we adopt this assumption and write an elementary operation as $U(\theta,\phi):=U\bigl(\theta,(\cos\phi, \sin\phi,0)\bigr)$.
	We also assume that the target operation that we attempt to perform has the form of $U(\theta,\phi)$.
	
	ORE is a typical error in NMR.
	When a qubit suffers from ORE, the unitary operation is deformed to
	\begin{equation}
		U^{(f)}(\theta,\phi):=\exp \bigl(-i \theta (\cos\phi \sigma_{x}+\sin \phi \sigma_{y}+f \sigma_{z} )/2\bigr)\sim U(\theta,\phi)-i f \sin(\theta/2) \sigma_{z}+ {\cal O}(f^2),
	\end{equation}
	where $f$ is an unknown small parameter that represents the magnitude of ORE.
	
	A (first-order) ORE-robust CP is defined as an operation sequence $U(\theta_{k},\phi_{k})\cdots U(\theta_{1},\phi_{1})$ consisting of $k$ elementary operations to compensate for the effect of ORE by satisfying the following condition:
	\begin{equation}
		U^{(f)}(\theta_{k},\phi_{k})\cdots U^{(f)}(\theta_{1},\phi_{1}) = U(\theta,\phi)+{\cal O}(f^{2})
		\label{eq:ore_robust}
	\end{equation}
	for the target operation $U(\theta,\phi)$.
	That is, the first-order ORE term of each elementary operation in a CP cancels each other out.
	We assume that $f$ is constant during the operations, which means that the timescale of the operation sequence is much shorter than the fluctuation timescale of $f$.
	We now focus on $k=3$ CPs, which have the smallest $k$ because no $k=2$ CPs exist \cite{bando2012concatenated}.

	\section{general form of $k=3$ ORE robust symmetric composite pulses}
	\label{sec:III}
	
	\subsection{variables and equations}
	
	We consider $k=3$ ORE robust CPs, $U(\theta_{3},\phi_{3})U(\theta_{2},\phi_{2})U(\theta_{1},\phi_{1})$.
	Hereinafter, we assume the time symmetry (mod $2 \pi$) of this sequence, that is, $\theta_{3}=\theta_{1}+2 n \pi$ and $\phi_{3}=\phi_{1}$.
	The reason why we take these assumptions is explained in Appendix \ref{appendix1}.
	To eliminate the maltivalency of variables, we parameterize $(\theta_{1},\theta_{2},\theta_{3})$ as
	\begin{align}
		\theta_{i}=\theta^{(p)}_{i}+2 n_{i} \pi,~~\theta^{(p)}_{1}=\theta^{(p)}_{3},~~0<\theta^{(p)}_{i}\leq 2 \pi,~~n_{i}\in {\mathbb N}~~(i=1,2,3),
	\end{align}
	where we use positivity of $\theta_{i}$'s and the time symmetry.
	
	We introduce the notations, $s_{i}$, $c_{i}$, and $U_{i}$, as follows:
	\begin{equation}
		s_{i}:=\sin(\theta^{(p)}_{i}/2) =(-1)^{n_{i}}\sin(\theta_{i}/2),~~c_{i}:=\cos(\theta^{(p)}_{i}/2) =(-1)^{n_{i}}\cos(\theta_{i}/2),~~
		U_{i}:=U(\theta^{(p)}_{i},\phi_{i})=(-1)^{n_{i}}U(\theta_{i},\phi_{i}).
	\end{equation}
	Note that $s_{i}$ is always positive and $U_{3}=U_{1}$.
	Using these variables, we rewrite the ORE robustness condition (\ref{eq:ore_robust}) as
	\begin{equation}
		(-1)^{n_{1}+n_{2}+n_{3}}\Bigl(s_{2}{\mathbb I}+s_{1}U_{1}U_{2}+s_{1}U^{\dagger}_{2}U^{\dagger}_{1}\Bigr)=0.
	\end{equation}
	Evidently, $n_{i}$'s do not affect this condition.
	Direct calculations show that the above equation is equivalent to
	\begin{equation}
		s_{2}+s_{1}(c_{2}c_{1}-\alpha s_{2}s_{1})=0,
		\label{eq:ORErobust}
	\end{equation}
	where $\alpha:=\cos(\phi_{2}-\phi_{1})$.
	
	A CP must reproduce the target operation $U(\theta,\phi)$ when $f=0$: $U(\theta_{3},\phi_{3})U(\theta_{2},\phi_{2})U(\theta_{1},\phi_{1})=U(\theta,\phi)$.
	In terms of $\theta^{(p)}_{i}$'s, this condition is rewritten as $U_{1}U_{2}U_{1}=(-1)^{n_{1}+n_{2}+n_{3}}U(\theta,\phi)$.
	We will evaluate the explicit form of this matrix equation.
	First, we note that a symmetric CP (mod $2 \pi$) can only implement a unitary operation in the form of $U(\theta,\phi)$, that is, the rotation direction of the target operation is in the $xy$ plane (Appendix \ref{appendix1}).
	We can easily verify this when $k=3$, and induction extends this result to general $k$ cases.
	The condition $U_{1}U_{2}U_{1}=(-1)^{n_{1}+n_{2}+n_{3}}U$ yields the following equations:
	\begin{align}
		&c_{2}(c^{2}_{1}-s^{2}_{1})-2\alpha c_{1}s_{1}s_{2}=(-1)^{n_{1}+n_{2}+n_{3}} c,\label{eq:target1}\\ 
		&2 s_{1}c_{1}c_{2}e^{-i (\phi_{2}-\phi_{1})}+c^{2}_{1}s_{2}-s^{2}_{1}s_{2}e^{-2i(\phi_{2}-\phi_{1})}=(-1)^{n_{1}+n_{2}+n_{3}} s e^{-i (\phi_{2}-\phi)}
		\label{eq:target2}
	\end{align}
	where $s:=\sin(\theta/2)$, $c:=\cos(\theta/2)$.
	We usually need to consider the case $0<\theta<2\pi$; thus, $s$ is positive.
	Eq. (\ref{eq:target1}) originates from the diagonal part of the condition, whereas Eq.~(\ref{eq:target2}) from the off-diagonal part.
	We obtain an ORE-robust sequence $(\theta^{(p)}_{1},\theta^{(p)}_{2},\phi_{1},\phi_{2})$ with arbitrary $(n_{1},n_{2},n_{3})$ by solving Eqs. (\ref{eq:ORErobust}), (\ref{eq:target1}), and (\ref{eq:target2}) simultaneously.
	We introduce the parameter $n=n_{1}+n_{2}+n_{3}\in {\mathbb N}$ because $n_{i}$'s appear only in this form.
	The equations above have solutions for any $n$, as shown below, and we can then choose $(n_{1},n_{2},n_{3})$ arbitrarily.
	Basically, smaller $n_{i}$ values are preferable for shortening the CP.
	Further note that, the freedom of $n_{i}$'s does not affect the first-order ORE robustness, which can be easily shown,
	but the higher-order ORE robustness is affected by the choice of $n_{i}$ in general.
	
	We treat the variables $(c_{1},c_{2},k:=\phi_{2}-\phi,l:=\phi_{2}-\phi_{1})$ instead of $(\theta^{(p)}_{1},\theta^{(p)}_{2},\phi_{1},\phi_{2})$.
	The variable $c_{i}$ has the one-to-one correspondence with $\theta^{(p)}_{i}$ because $0<\theta^{(p)}_{i}/2 \leq \pi$ and $s_{i}$ is always written as $s_{i}=\sqrt{1-c^{2}_{i}}$ without sign ambiguity.
	For the same reason as above, we use $c$ as a parameter instead of $\theta$ because we now consider $0<\theta \leq 2 \pi$.
	
	We now discuss how to solve Eqs.~(\ref{eq:ORErobust}), (\ref{eq:target1}), and (\ref{eq:target2}).
	Eq. (\ref{eq:target2}) apparently leads to two (real and imaginary) equations, but it turns out to be one equation when Eq. (\ref{eq:target1}) is satisfied (see Appendix B).
	Therefore, Eqs. (\ref{eq:ORErobust}), (\ref{eq:target1}), and (\ref{eq:target2}) lead to three equations for four variables of $(c_1, c_2, k, l)$: one can determine only three variables.
	We take $c_1$ as a free parameter and solve for $(c_2, k, l)$.
	To solve the equations, we first focus on Eqs. (\ref{eq:ORErobust}) and (\ref{eq:target1}).	
	They do not contain $k$, and thus lead to $c_2$ and $l$ as functions 
	of $c_1$ and $c$.
	Then, Eq. (\ref{eq:target2}) determines $k$ as a function of $c_1$ and $c$. 
	Note that $c_1$ is not a perfectly free parameter, but it exists in a certain region determined by $c$:
	we will discuss it in the next section.
	
	Note that none of the variables $(c_1, c_2, k, l)$ depend on $\phi$.
	Accordingly, the original variables are given as
	\begin{equation}
		\theta^{(p)}_{1},~~\theta^{(p)}_{2}(\theta,\theta^{(p)}_{1}),~~\phi_{1}=\phi+k(\theta,\theta^{(p)}_{1})-l(\theta,\theta^{(p)}_{1}),~~\phi_{2}=\phi+k(\theta,\theta^{(p)}_{1}),
	\end{equation}
	where we use the one-to-one correspondence between $c_{i}$ ($c$) and $\theta^{(p)}_{i}$ ($\theta$).
	Intrinsically, we only need the information of $\theta$ of the target operation $U(\theta,\phi)$ to solve the equations.
	First, we obtain $(\theta^{(p)}_{1},\theta^{(p)}_{2},\phi_{1},\phi_{2})$ for the target operation $U(\theta,0)$ by taking $\phi=0$.
	Then, we obtain the solutions for the target operation $U(\theta,\phi)$ by transforming $(\theta^{(p)}_{1},\theta^{(p)}_{2},\phi_{1},\phi_{2})$ to $(\theta^{(p)}_{1},\theta^{(p)}_{2},\phi_{1}+\phi,\phi_{2}+\phi)$.
	This behaviour is a reflection of the rotational symmetry around the $z$ direction.

	\subsection{solutions}
	
	Eqs. (\ref{eq:ORErobust}) and (\ref{eq:target1}) provide two types of solutions, depending on the parity of $n$.
	$c_{2}$ ($s_{2}$) and $\alpha$ are determined as a function of $c_{1}$ and $c$ ($\theta$):
	\begin{align}
		c_{2}=&~-(-1)^{n}c s^{2}_{1}-c_{1}\sqrt{1-c^{2}s^{2}_{1}},~~s_{2}=s_{1}(\sqrt{1-c^{2}s^{2}_{1}}-(-1)^{n}c c_{1}) \label{eq:solution1}\\
		\alpha=&~1-\frac{\sqrt{1-c^{2}s^{2}_{1}}+(-1)^{n}c c_{1}}{2 s^{2}_{1} \sqrt{1-c^{2}s^{2}_{1}}-(-1)^{n}c c_{1}} .\label{eq:solution2}
	\end{align}
	where $s_{1}$ is uniquely determined by $s_{1}=\sqrt{1-c^{2}_{1}}$.
	See Appendix \ref{appendix3} for the derivation.
	Because $c_{1}$ remains a free parameter as mentioned above, we have an {\it infinite} number of choices for the construction of $U(\theta,\phi)$ in an ORE-robust manner even in the $k=3$ case.
	
\begin{figure}[h]
	\begin{center}
		\includegraphics[width=180mm]{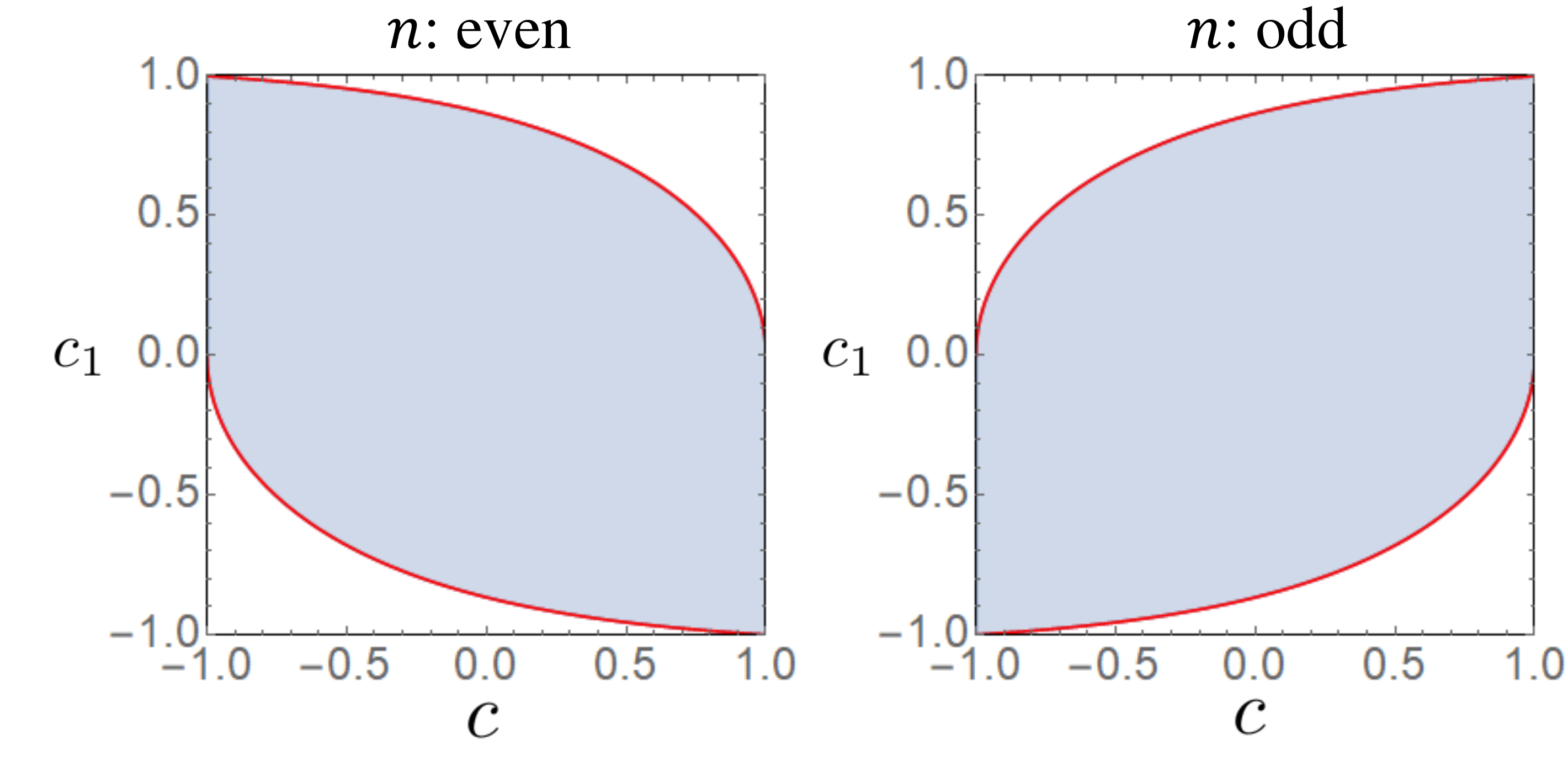}
		\caption{The region in which $c_{1}$ can be taken to construct an ORE-robust CP. The blue shaded region is the allowed region, whereas the red solid lines are the upper bound $c_{1,n,+}$ and lower bound $c_{1,n,-}$.
		\label{fig:existence}
		}
	\end{center}
	\end{figure}

	Although the number of choices for $c_{1}$ is infinite, there are upper and lower bounds of $c_{1}$ for a fixed $c$.
	Note that $\alpha=\cos(\phi_{2}-\phi_{1})$ must be $-1 \leq \alpha \leq 1$.
	Equation (\ref{eq:solution2}) guarantees only $\alpha \leq 1$ but not $\alpha\geq -1$.
	The latter inequality gives additional upper and lower bounds of $c_{1}$ as
	\begin{equation}
		c_{1,n,-}:=-\frac{\sqrt{3-c^{2}+(-1)^{n}c\sqrt{3+c^{2}}}}{2}\leq c_{1} \leq \frac{\sqrt{3-c^{2}-(-1)^{n}c\sqrt{3+c^{2}}}}{2}=:c_{1,n,+}.
		\label{eq:bound}
	\end{equation}
	Appendix \ref{appendix3} also shows the detailed calculations for this.
	The region in which $c_{1}$ takes is shown in Fig. \ref{fig:existence}.
	The regions of odd and even parity are a mirror image of each other.
	This is because the index $n$ in the solution appears only in the form of $(-1)^{n}c$.
	
	Here, we show a ``recipe" to construct an ORE-robust CP:
	\begin{enumerate}
		\item Fix $\theta$ of the target operation that we perform.
		\item Fix $n_{i}$'s arbitrarily. (Smaller numbers are basically preferred in order to shorten the operation time.)
		\item Fix $c_{1}$ ($\theta^{(p)}_{1}$) arbitrarily in the region shown by Eq.~(\ref{eq:bound}).
		\item $c_{2}$ and $\alpha=\cos(\phi_{2}-\phi_{1})$ are automatically determined by Eqs.~(\ref{eq:solution1}) and (\ref{eq:solution2}).
		\item Solve $k=\phi-\phi_{2}$ to satisfy Eq. (\ref{eq:target2}).
		\item Decide $\phi$ of the target operation and then $\phi_{1}$ ($\phi_{2}$) is automatically determined as $\phi_{1}=\phi+k-l$ ($\phi_{2}=\phi+k$).
	\end{enumerate}
	This method exhausts all possibilities of the $k=3$ symmetric ORE-robust CPs.

	\subsection{relation with CORPSE}
	
	We discuss how the above general construction reproduces the CORPSE family, a well-known family of ORE-robust symmetric CPs with the sequence  $U(\theta^{(c)}_{3},\phi^{(c)}_{3})U(\theta^{(c)}_{2},\phi^{(c)}_{2})U(\theta^{(c)}_{1},\phi^{(c)}_{1})$. The parameters are written as
	\begin{align}
		\theta^{(c)}_{1}&=\theta/2-\kappa+2 \nu_{1}\pi,\nonumber\\
		\theta^{(c)}_{2}&=2 \nu_{2}\pi-2 \kappa,\nonumber\\
		\theta^{(c)}_{3}&=\theta/2-\kappa+2 \nu_{3}\pi,\nonumber\\
		\phi^{(c)}_{1}&=\phi^{(c)}_{3}=\phi,\nonumber\\
		\phi^{(c)}_{2}&=\phi+\pi,
		\label{eq:corpse}
	\end{align}
	where $\kappa=\arcsin[\sin(\theta/2)/2]$, and $U(\theta,\phi)$ is the target operation.
	We first consider short CORPSE, which is a member of the CORPSE family with the parameters $\nu_{1}=0$, $\nu_{2}=1$, and $\nu_{3}=0$.
	This corresponds to $\theta_{i}\leq 2\pi$ $(i=1,2,3)$ in our general expression, or equivalently, $n_{i}=0$ $(i=1,2,3)$.
	Note that this short CORPSE performs $U(2\pi-\theta,\phi+\pi)$ instead of $U(\theta,\phi)$ although the only difference between them is the global phase $-1$.
	The actual sequence of short CORPSE that performs $U(\theta,\phi)$ is obtained by transforming $(\theta,\phi)$ to $(2\pi-\theta,\phi+\pi)$:
	\begin{align}
		\theta^{\rm (sc)}_{1}&=\theta^{\rm (sc)}_{3}=\pi-\theta/2-\kappa,\nonumber\\
		\theta^{\rm (sc)}_{2}&=2 \pi-2 \kappa,\nonumber\\
		\phi^{\rm (sc)}_{1}&=\phi^{\rm (sc)}_{3}=\phi+\pi,\nonumber\\
		\phi^{\rm (sc)}_{2}&=\phi,
		\label{eq:short_corpse}
	\end{align}
	where $\kappa$ is the same because $\sin \bigl((2\pi-\theta)/2\bigr)=\sin(\theta/2)$.
	
	To check the relation between our description and short CORPSE,
	we calculate $\cos^{2}(\theta^{\rm (sc)}_{1}/2)$ as
	\begin{align}
		\cos^{2}(\theta^{\rm (sc)}_{1}/2)=&\frac{\cos(\theta^{\rm (sc)}_{1})+1}{2}
		=\frac{\cos(\pi-\theta/2-\kappa)+1}{2}=\frac{-\cos(\theta/2+\kappa)+1}{2}\nonumber\\
		=&\frac{-c \sqrt{1-(s/2)^{2}}+s^{2}/2 +1}{2}= \frac{3-c^{2}-c \sqrt{3+c^{2}}}{4},
	\end{align}
	where we use $\cos \bigl(\arcsin(a)\bigr)=\sqrt{1-a^{2}}$.
	To determine the sign of $\cos(\theta^{\rm (sc)}_{1}/2)$, we differentiate $\theta^{\rm (sc)}_{1}$ in Eq.~(\ref{eq:short_corpse}), with respect to $\theta$:
	\begin{equation}
		\frac{d \theta^{\rm (sc)}_{1}}{d \theta}=\frac{d}{d \theta}\Bigl(\pi-\frac{\theta}{2}-\kappa\Bigr)=-\frac{1}{2}\Bigl(1+\frac{c}{\sqrt{3+c^{2}}}\Bigr).
	\end{equation}
	This is always negative for $0<\theta<2\pi$, which means that $\cos(\theta^{\rm (sc)}_{1}/2)$ is a monotonically increasing function of $\theta$.
	As $\cos(\theta^{\rm (sc)}_{1}/2)=0$ when $\theta=0$, $\cos(\theta^{\rm (sc)}_{1}/2)$ is always positive.
	Thus, we find
	\begin{equation}
		\cos(\theta^{\rm (sc)}_{1}/2)=\frac{\sqrt{3-c^{2}-c \sqrt{3+c^{2}}}}{2}.
	\end{equation}
	This is the same as $c_{1,0,+}$ with $n=0$ in Eq. (\ref{eq:bound}).
	Similarly, other variables $\theta^{\rm (sc)}_{2}$, $\phi^{\rm (sc)}_{1}$ and $\phi^{\rm (sc)}_{2}$ are easily proved to be equal to our variables $\theta^{(p)}_{2}$, $\phi_{1}$, and $\phi_{2}$, respectively, when $c_{1}=c_{1,0,+}$.
	Thus, we found that our CP with $(n_{1},n_{2},n_{3})=(0,0,0)$ and $c_{1}=c_{1,0,+}$ is equivalent to short CORPSE.
	
	For other values of $n_{i}$'s, we find that the choice of $c_{1}=c_{1,n,+}$ reproduces the CPs in the CORPSE family.
	Particularly, the case of $(n_{1},n_{2},n_{3})=(1,0,0)$ corresponds to a CORPSE sequence with $(\nu_{1},\nu_{2},\nu_{3})=(1,1,0)$.
	When comparing our construction with the CORPSE family, we should be careful that the CORPSE family sometimes implement not $U(\theta,\phi)$ but $U(2\pi-\theta,\phi+\pi)$, depending on $\nu_{i}$'s, such as short CORPSE.
	
	The starting point for constructing the CORPSE family is to set the angle between $\phi_{1}$ and $\phi_{2}$ to $\cos(\phi_{2}-\phi_{1})=-1$.
	One may notice that the bottom edge of the region ($c_{1}=c_{1,0,-}$) has a similar behaviour to the CORPSE family:
	$\cos(\phi_{2}-\phi_{1})=-1$ on this edge as in the case of the top edge $c_{1}=c_{1,0,+}$ corresponding to the CORPSE family.
	This edge  ($c_{1}=c_{1,0,-}$), however, corresponds not to the CORPSE family but to its ``twin".
	Generally, there are two choices of $c_{1}$ (or $\theta_{1}$) for a fixed $\alpha$.
	See Appendix \ref{appendix4} for the detail.

	\section{Performance evaluation}
	\label{sec:IIII}
	
	\subsection{gate infidelity}
	
	Here, we evaluate the accuracy of the ORE-robust CPs found above. 
	We use the gate infidelity as a measure of their accuracy.
	The gate infidelity of an operation is defined as
	\begin{equation}
		F:=1-\frac{|{\rm tr}(U^{\dagger}U')|}{2},
	\end{equation}
	where $U'$ is an operation with ORE and $U$ is the corresponding errorless gate.
	The gate infidelity equals $0$ if and only if $U=U'$, and a smaller gate infidelity implies a better accuracy.
	It should be noted that the gate infidelity behaves as $F\approx {\cal O}(f^{4})$ for ORE-robust CPs \cite{JONES201191}.
	Appendix \ref{appendix5} provides that a simple proof of this behaviour for more general error models.
	
	\begin{figure}[h]
	\begin{center}
		\includegraphics[width=180mm]{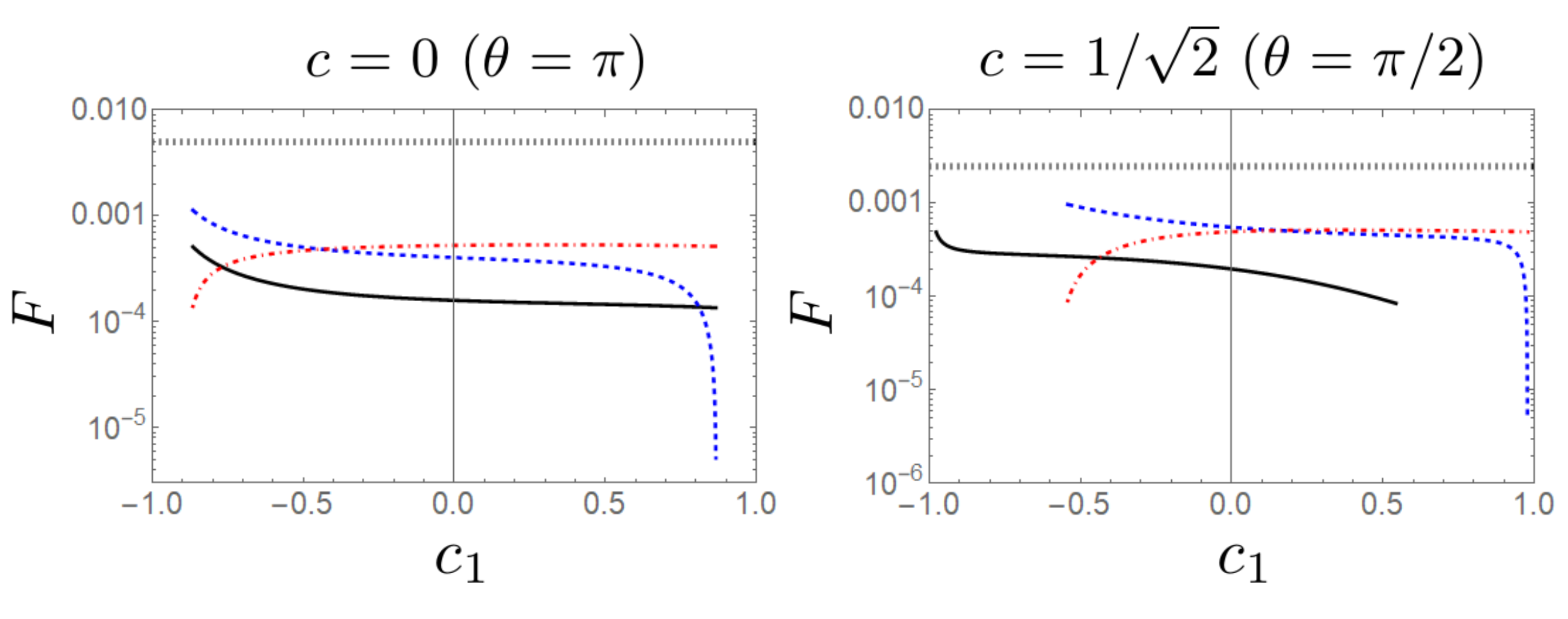}
		\caption{Gate infidelity for the cases of $c=1/\sqrt{2}~(\theta=\pi/2)$ and $c=0 ~(\theta=\pi)$.
			The solid, dashed, and dot-dashed curves correspond to $(n_{1},n_{2},n_{3})=(0,0,0)$, $(n_{1},n_{2},n_{3})=(1,0,0)$, and $(n_{1},n_{2},n_{3})=(0,1,0)$, respectively.
			The dotted straight line represents the gate infidelity of the elementary operation with ORE.
			The error parameter is set to $f=0.1$.
			All curves for $c=0$ have the same existence range because $c_{1,n,\pm}=\pm \sqrt{3}/2$, regardless of $n$ in this case.
			On the other hand, the curve corresponding to $(n_{1},n_{2},n_{3})=(0,0,0)$ for the case of $c=1/\sqrt{2}$ has a different existence range than the other curves: $c_{1,n,\pm}$ depend on $n$ unless $c=0$.
			\label{fig:fidelity}}
	\end{center}
	\end{figure}

	Fig. \ref{fig:fidelity} plots the gate infidelity with several combinations of $n_{i}$'s for the cases of $c=1/\sqrt{2}~(\theta=\pi/2)$ and $c=0 ~(\theta=\pi)$. 
	Note that specific values of $n$ and $n_{i}$'s can affect the gate infidelity at a higher order of $f$, whereas it does not affect the first order ORE robustness.
	This implies that the gate infidelity could be improved when we consider a larger $n$ although we do not evaluate such cases in this paper.
	In both cases of $c=0$ and $1/\sqrt{2}$, the curves corresponding to $(n_{1},n_{2},n_{3})=(0,0,0)$ and $(n_{1},n_{2},n_{3})=(1,0,0)$ have the minimum at the right end $c_{1}=c_{1,n,+}$, whereas the curve for $(n_{1},n_{2},n_{3})=(0,1,0)$ has the minimum at the left end $c_{1}=c_{1,n,-}$.
	These points correspond to the CPs in the CORPSE family ($c_{1}=c_{1,n,+}$) or its twin family ($c_{1}=c_{1,n,-}$), which are characterized by $\cos(\phi_{2}-\phi_{1})=-1$.
	Moreover, the right end of the curve for $(n_{1},n_{2},n_{3})=(1,0,0)$ shows a good accuracy; it is approximately in the order of $f^{6}$.
	This point corresponds to a CP in the CORPSE family with $(\nu_{1},\nu_{2},\nu_{3})=(1,1,0)$ in Eq. (\ref{eq:corpse}).
	(Hereinafter, we call this case ``fundamental CORPSE''.)
	The fourth-order term of the gate infidelity for fundamental CORPSE {\it does not} vanish but has a small value.
	
	However, the CORPSE family does not necessarily have the best accuracy when we consider a measure other than the gate infidelity.
	For example, let us consider the state fidelity $|\langle \psi|U^{\dagger}U'|\psi\rangle|$ when we take the eigenvector of $\sigma_{z}$ corresponding to eigenvalue $1$ as $|\psi\rangle$ (the initial state of the dynamics).
	In this case, we can find a CP that has a better accuracy than the CORPSE family. See Appendix~\ref{appendix6}.
	
	The gate infidelity corresponds to the worst value among all the state infidelities for all initial states (Appendix \ref{appendix5}).
	Thus, if we apply an ORE-robust CP to an unknown initial state, we can conclude that the CORPSE family or its twin has the best accuracy, at least when $c=0,1/\sqrt{2}$.
	Particularly, fundamental CORPSE appears to provide the best performance compared with other choices of $c_{1}$.

	\subsection{total operation time}
	
	Another measure for performance of a quantum operation is time required for the operation.
	To avoid decoherence and other undesirable environmental effects, a short operation time (or operation length) is preferred.
	We assume that the strength of the control field is constant $\omega$ throughout the operations.
	The parameter $\theta_{i}$ is given as $\theta_{i}=\omega t_{i}$, where $t_{i}$ is the time required for each operation.
	In this terminology, the operation time of CPs is simply proportional to $L=\theta_{1}+\theta_{2}+\theta_{3}=2\theta^{(p)}_{1}+\theta^{(p)}_{2}+2 \pi n$.
	We evaluate the operation time $L$ as a function of $(c,c_{1},n)$ because the other parameters are uniquely determined by these variables.
	The phase $\phi$ of the target operation does not affect $L$.
	
	\begin{figure}[h]
		\begin{center}
		\includegraphics[width=175mm]{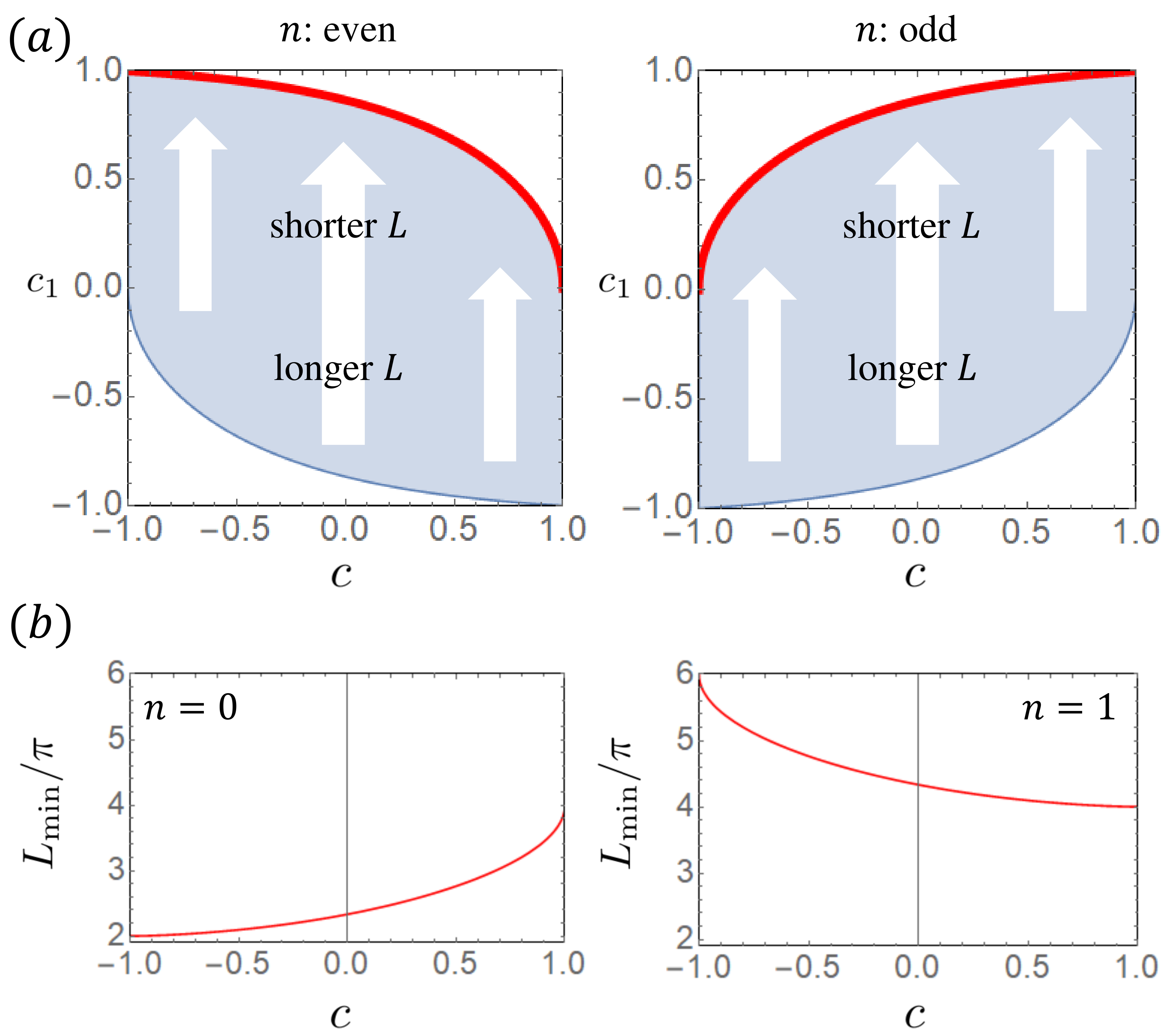}
		\caption{Shortest operation time. (a) The existence range of $c_1$ as a function of $c$. 
			$c_{1} =c_{1,n,+}$ on the solid thick line provides the minimum operation time $L_{\rm min}(c,n)$ for either case of even $n$ (the left panel) or odd $n$ (the right panel). 
			(b) The minimum operation time $L_{\rm min} (c,0)$ (the left panel) and $L_{\rm min} (c,1)$
			(the right panel). These lines correspond to the solid thick (red) lines in (a). 
			\label{fig:length_c}
			}
		\end{center}
	\end{figure}
	
	First, we consider the freedom of $c_{1}$ for fixed $n$ and $c$.
	The function $L$ is explicitly given as
	\begin{equation}
		L(c,c_{1},n)=4 \arccos (c_{1})+2 \arccos \bigl( c_{2} (c,c_{1})\bigr)+2 \pi n,
	\end{equation}
	where we explicitly show the $c_{1}$ dependence of $c_{2}$ and use $c_{1}=\cos(\theta_{1}/2)$ and $c=\cos(\theta/2)$.
	The derivative of $L$ with respect to $c_{1}$ is given as
	\begin{equation}
		\frac{\partial L(c,c_{1},n)}{\partial c_{1}}=-\frac{2}{\sqrt{1-c^{2}_{1}}}\biggl( 1+(-1)^{n}\frac{c c_{1}}{\sqrt{1-(1-c^{2}_{1})c^{2}}}\biggr),
		\label{eq:partial_negative}
	\end{equation}
	which is always negative regardless of $n$ and $c$.
	This implies that $L$ is a monotonically decreasing function of $c_{1}$.
	Thus, we found that the largest $c_{1}=c_{1,n,+}$ provides the shortest time $L(c,c_{1,n,+},n\bigr)$ for any fixed value of $c$ and $n$. See Fig. \ref{fig:length_c}~(a).
	We define The minimum operation time for a fixed $n$ as $L_{\rm min}(c,n):=L\bigl(c,c_{1,n,+},n\bigr)$.

	We then consider which of $L_{\rm min}(c,0)$ or $L_{\rm  min}(c,1)$ is shorter for a given $c$.
	We trivially have the sequences $L_{\rm min}(c,0)< L_{\rm min}(c,2)< L_{\rm min}(c,4)<\cdots$ and $L_{\rm min}(c,1)< L_{\rm min}(c,3)< L_{\rm min}(c,5)<\cdots$.
	However, it should be verified whether $L_{\rm min}(c,0)\leq L_{\rm min}(c,1)$ or $L_{\rm min}(c,1)\leq L_{\rm min}(c,0)$ because the parity difference between odd and even $n$ yields the different solutions.
	Fig. \ref{fig:length_c}~(b) shows $L(c,c_{1,0,+},0)$ and $L(c,c_{1,1,+},1)$ and evidently implies $L(c,c_{1,0,+},0)\leq L(c,c_{1,1,+},1)$.
	
	To mathematically justify this result, we calculate the derivative of $L_{\rm min}(c,n)$ with respect to $c$ and obtain
	\begin{equation}
		\frac{d L_{\rm min}(c,n)}{d c}=(-1)^{n}\frac{2\gamma \Bigl(\sqrt{1-(1-c^{2}_{1,n,+})c^{2  }}+(-1)^{n}c c_{1,n,+}\Bigr)+1-c^{2}_{1,n,+}}{\sqrt{(1-c^{2}_{1,n,+})(1-(1-c^{2}_{1,n,+})c^{2})}}
		\label{eq:total_negative}
	\end{equation}
	where $\gamma=-(-1)^{n}d c_{1,n,+}/d c$ and it is always positive.
	The numerator is easily shown to be positive.
	Thus, we find that $L_{\rm min}(c,n)$ is a monotonically increasing (decreasing) function of $c$ when $n$ is $0$ ($1$).
	$L_{\rm min}(c,0)$ has the maximal value of $4\pi$ at $c=1$, whereas $L_{\rm min}(c,1)$ has the minimal value of $4\pi$ at $c=1$ as in Fig. \ref{fig:length_c}~(b).
	Thus, $L_{\rm min}(c,0)$ gives the shortest time for any values of $c$.
	In other word, short CORPSE provides the shortest operation time among all the $k=3$ symmetric ORE-robust CPs.
		
	\section{Conclusion}
	\label{sec:IIIII}
	
	In this study, we provided a systematic construction of ORE-robust CPs comprising three elementary operations.
	We assumed that the operation sequence is symmetric modulo $2\pi$.
	Under this condition, we found general solutions for an ORE-robust CP implementing an arbitrary $\theta$-rotation.
	We have one continuous degree of freedom, even after solving all the conditions for the ORE-robust $\theta$-rotation.
	This implies that there are infinitely many $k=3$ symmetric CPs that implement an arbitrary $\theta$-rotation.
	This sequence includes the CORPSE family, a well-known and well-investigated family of ORE-robust CPs.
	We then evaluated the performance of the CPs in terms of the gate infidelity and the total operation time.
	In our calculation, the CORPSE family provides the best accuracy among all the choices evaluated, at least when $\theta=\pi$ and $\pi/2$.
	We also proved that short CORPSE, one of the CPs in the CORPSE family, has the shortest operation time among the symmetric ORE-robust CPs with three elementary operations.
	
	So far, the performance of the CORPSE family has never been investigated in comparison to other ORE-robust CPs.
	Our construction, which provides a wide class of ORE-robust CPs, including the CORPSE family, enabled us to validate the performance of the CORPSE family among this wide class of ORE-robust CPs.
	Our results show that fundamental CORPSE and short CORPSE (or their twins) actually have good performance in terms of gate infidelity and total operation time compared among this wide class of ORE-robust CPs.
	
	\bibliography{OREpaper2}

\begin{thebibliography}{41}%
\makeatletter
\providecommand \@ifxundefined [1]{%
 \@ifx{#1\undefined}
}%
\providecommand \@ifnum [1]{%
 \ifnum #1\expandafter \@firstoftwo
 \else \expandafter \@secondoftwo
 \fi
}%
\providecommand \@ifx [1]{%
 \ifx #1\expandafter \@firstoftwo
 \else \expandafter \@secondoftwo
 \fi
}%
\providecommand \natexlab [1]{#1}%
\providecommand \enquote  [1]{``#1''}%
\providecommand \bibnamefont  [1]{#1}%
\providecommand \bibfnamefont [1]{#1}%
\providecommand \citenamefont [1]{#1}%
\providecommand \href@noop [0]{\@secondoftwo}%
\providecommand \href [0]{\begingroup \@sanitize@url \@href}%
\providecommand \@href[1]{\@@startlink{#1}\@@href}%
\providecommand \@@href[1]{\endgroup#1\@@endlink}%
\providecommand \@sanitize@url [0]{\catcode `\\12\catcode `\$12\catcode
  `\&12\catcode `\#12\catcode `\^12\catcode `\_12\catcode `\%12\relax}%
\providecommand \@@startlink[1]{}%
\providecommand \@@endlink[0]{}%
\providecommand \url  [0]{\begingroup\@sanitize@url \@url }%
\providecommand \@url [1]{\endgroup\@href {#1}{\urlprefix }}%
\providecommand \urlprefix  [0]{URL }%
\providecommand \Eprint [0]{\href }%
\providecommand \doibase [0]{https://doi.org/}%
\providecommand \selectlanguage [0]{\@gobble}%
\providecommand \bibinfo  [0]{\@secondoftwo}%
\providecommand \bibfield  [0]{\@secondoftwo}%
\providecommand \translation [1]{[#1]}%
\providecommand \BibitemOpen [0]{}%
\providecommand \bibitemStop [0]{}%
\providecommand \bibitemNoStop [0]{.\EOS\space}%
\providecommand \EOS [0]{\spacefactor3000\relax}%
\providecommand \BibitemShut  [1]{\csname bibitem#1\endcsname}%
\let\auto@bib@innerbib\@empty
\bibitem [{\citenamefont {Nielsen}\ and\ \citenamefont
  {Chuang}(2000)}]{Nielsen2000}%
  \BibitemOpen
  \bibfield  {author} {\bibinfo {author} {\bibfnamefont {M.}~\bibnamefont
  {Nielsen}}\ and\ \bibinfo {author} {\bibfnamefont {I.}~\bibnamefont
  {Chuang}},\ }\href {https://books.google.co.jp/books?id=aai-P4V9GJ8C} {\emph
  {\bibinfo {title} {Quantum Computation and Quantum Information}}},\ Cambridge
  Series on Information and the Natural Sciences\ (\bibinfo  {publisher}
  {Cambridge University Press},\ \bibinfo {year} {2000})\BibitemShut {NoStop}%
\bibitem [{\citenamefont {Bennett}\ and\ \citenamefont
  {DiVincenzo}(2000)}]{bennett2000quantum}%
  \BibitemOpen
  \bibfield  {author} {\bibinfo {author} {\bibfnamefont {C.~H.}\ \bibnamefont
  {Bennett}}\ and\ \bibinfo {author} {\bibfnamefont {D.~P.}\ \bibnamefont
  {DiVincenzo}},\ }\bibfield  {title} {\bibinfo {title} {Quantum information
  and computation},\ }\href@noop {} {\bibfield  {journal} {\bibinfo  {journal}
  {nature}\ }\textbf {\bibinfo {volume} {404}},\ \bibinfo {pages} {247}
  (\bibinfo {year} {2000})}\BibitemShut {NoStop}%
\bibitem [{\citenamefont {Nakahara}(2008)}]{nakahara2008quantum}%
  \BibitemOpen
  \bibfield  {author} {\bibinfo {author} {\bibfnamefont {M.}~\bibnamefont
  {Nakahara}},\ }\href@noop {} {\emph {\bibinfo {title} {Quantum computing:
  from linear algebra to physical realizations}}}\ (\bibinfo  {publisher} {CRC
  press},\ \bibinfo {year} {2008})\BibitemShut {NoStop}%
\bibitem [{\citenamefont {Ekert}(1991)}]{ekert1991quantum}%
  \BibitemOpen
  \bibfield  {author} {\bibinfo {author} {\bibfnamefont {A.~K.}\ \bibnamefont
  {Ekert}},\ }\bibfield  {title} {\bibinfo {title} {Quantum cryptography based
  on bell’s theorem},\ }\href@noop {} {\bibfield  {journal} {\bibinfo
  {journal} {Physical review letters}\ }\textbf {\bibinfo {volume} {67}},\
  \bibinfo {pages} {661} (\bibinfo {year} {1991})}\BibitemShut {NoStop}%
\bibitem [{\citenamefont {Gisin}\ and\ \citenamefont
  {Thew}(2007)}]{gisin2007quantum}%
  \BibitemOpen
  \bibfield  {author} {\bibinfo {author} {\bibfnamefont {N.}~\bibnamefont
  {Gisin}}\ and\ \bibinfo {author} {\bibfnamefont {R.}~\bibnamefont {Thew}},\
  }\bibfield  {title} {\bibinfo {title} {Quantum communication},\ }\href@noop
  {} {\bibfield  {journal} {\bibinfo  {journal} {Nature photonics}\ }\textbf
  {\bibinfo {volume} {1}},\ \bibinfo {pages} {165} (\bibinfo {year}
  {2007})}\BibitemShut {NoStop}%
\bibitem [{\citenamefont {Chen}\ \emph {et~al.}(2021)\citenamefont {Chen},
  \citenamefont {Zhang}, \citenamefont {Chen}, \citenamefont {Cai},
  \citenamefont {Liao}, \citenamefont {Zhang}, \citenamefont {Chen},
  \citenamefont {Yin}, \citenamefont {Ren}, \citenamefont {Chen} \emph
  {et~al.}}]{chen2021integrated}%
  \BibitemOpen
  \bibfield  {author} {\bibinfo {author} {\bibfnamefont {Y.-A.}\ \bibnamefont
  {Chen}}, \bibinfo {author} {\bibfnamefont {Q.}~\bibnamefont {Zhang}},
  \bibinfo {author} {\bibfnamefont {T.-Y.}\ \bibnamefont {Chen}}, \bibinfo
  {author} {\bibfnamefont {W.-Q.}\ \bibnamefont {Cai}}, \bibinfo {author}
  {\bibfnamefont {S.-K.}\ \bibnamefont {Liao}}, \bibinfo {author}
  {\bibfnamefont {J.}~\bibnamefont {Zhang}}, \bibinfo {author} {\bibfnamefont
  {K.}~\bibnamefont {Chen}}, \bibinfo {author} {\bibfnamefont {J.}~\bibnamefont
  {Yin}}, \bibinfo {author} {\bibfnamefont {J.-G.}\ \bibnamefont {Ren}},
  \bibinfo {author} {\bibfnamefont {Z.}~\bibnamefont {Chen}}, \emph {et~al.},\
  }\bibfield  {title} {\bibinfo {title} {An integrated space-to-ground quantum
  communication network over 4,600 kilometres},\ }\href@noop {} {\bibfield
  {journal} {\bibinfo  {journal} {Nature}\ }\textbf {\bibinfo {volume} {589}},\
  \bibinfo {pages} {214} (\bibinfo {year} {2021})}\BibitemShut {NoStop}%
\bibitem [{\citenamefont {Helstrom}(1976)}]{helstrom1976quantum}%
  \BibitemOpen
  \bibfield  {author} {\bibinfo {author} {\bibfnamefont {C.~W.}\ \bibnamefont
  {Helstrom}},\ }\href@noop {} {\emph {\bibinfo {title} {Quantum detection and
  estimation theory}}},\ Vol.~\bibinfo {volume} {84}\ (\bibinfo  {publisher}
  {Academic press New York},\ \bibinfo {year} {1976})\BibitemShut {NoStop}%
\bibitem [{\citenamefont {Caves}(1981)}]{caves1981quantum}%
  \BibitemOpen
  \bibfield  {author} {\bibinfo {author} {\bibfnamefont {C.~M.}\ \bibnamefont
  {Caves}},\ }\bibfield  {title} {\bibinfo {title} {Quantum-mechanical noise in
  an interferometer},\ }\href@noop {} {\bibfield  {journal} {\bibinfo
  {journal} {Physical Review D}\ }\textbf {\bibinfo {volume} {23}},\ \bibinfo
  {pages} {1693} (\bibinfo {year} {1981})}\BibitemShut {NoStop}%
\bibitem [{\citenamefont {Holevo}(2011)}]{holevo2011probabilistic}%
  \BibitemOpen
  \bibfield  {author} {\bibinfo {author} {\bibfnamefont {A.~S.}\ \bibnamefont
  {Holevo}},\ }\href@noop {} {\emph {\bibinfo {title} {Probabilistic and
  statistical aspects of quantum theory}}},\ Vol.~\bibinfo {volume} {1}\
  (\bibinfo  {publisher} {Springer Science \& Business Media},\ \bibinfo {year}
  {2011})\BibitemShut {NoStop}%
\bibitem [{\citenamefont {Vandersypen}\ and\ \citenamefont
  {Chuang}(2005)}]{RevModPhys.76.1037}%
  \BibitemOpen
  \bibfield  {author} {\bibinfo {author} {\bibfnamefont {L.~M.~K.}\
  \bibnamefont {Vandersypen}}\ and\ \bibinfo {author} {\bibfnamefont {I.~L.}\
  \bibnamefont {Chuang}},\ }\bibfield  {title} {\bibinfo {title} {Nmr
  techniques for quantum control and computation},\ }\href
  {https://doi.org/10.1103/RevModPhys.76.1037} {\bibfield  {journal} {\bibinfo
  {journal} {Rev. Mod. Phys.}\ }\textbf {\bibinfo {volume} {76}},\ \bibinfo
  {pages} {1037} (\bibinfo {year} {2005})}\BibitemShut {NoStop}%
\bibitem [{\citenamefont {Rahman}\ \emph {et~al.}(2007)\citenamefont {Rahman},
  \citenamefont {Wellard}, \citenamefont {Bradbury}, \citenamefont {Prada},
  \citenamefont {Cole}, \citenamefont {Klimeck},\ and\ \citenamefont
  {Hollenberg}}]{PhysRevLett.99.036403}%
  \BibitemOpen
  \bibfield  {author} {\bibinfo {author} {\bibfnamefont {R.}~\bibnamefont
  {Rahman}}, \bibinfo {author} {\bibfnamefont {C.~J.}\ \bibnamefont {Wellard}},
  \bibinfo {author} {\bibfnamefont {F.~R.}\ \bibnamefont {Bradbury}}, \bibinfo
  {author} {\bibfnamefont {M.}~\bibnamefont {Prada}}, \bibinfo {author}
  {\bibfnamefont {J.~H.}\ \bibnamefont {Cole}}, \bibinfo {author}
  {\bibfnamefont {G.}~\bibnamefont {Klimeck}},\ and\ \bibinfo {author}
  {\bibfnamefont {L.~C.~L.}\ \bibnamefont {Hollenberg}},\ }\bibfield  {title}
  {\bibinfo {title} {High precision quantum control of single donor spins in
  silicon},\ }\href {https://doi.org/10.1103/PhysRevLett.99.036403} {\bibfield
  {journal} {\bibinfo  {journal} {Phys. Rev. Lett.}\ }\textbf {\bibinfo
  {volume} {99}},\ \bibinfo {pages} {036403} (\bibinfo {year}
  {2007})}\BibitemShut {NoStop}%
\bibitem [{\citenamefont {Bason}\ \emph {et~al.}(2012)\citenamefont {Bason},
  \citenamefont {Viteau}, \citenamefont {Malossi}, \citenamefont {Huillery},
  \citenamefont {Arimondo}, \citenamefont {Ciampini}, \citenamefont {Fazio},
  \citenamefont {Giovannetti}, \citenamefont {Mannella},\ and\ \citenamefont
  {Morsch}}]{bason2012high}%
  \BibitemOpen
  \bibfield  {author} {\bibinfo {author} {\bibfnamefont {M.~G.}\ \bibnamefont
  {Bason}}, \bibinfo {author} {\bibfnamefont {M.}~\bibnamefont {Viteau}},
  \bibinfo {author} {\bibfnamefont {N.}~\bibnamefont {Malossi}}, \bibinfo
  {author} {\bibfnamefont {P.}~\bibnamefont {Huillery}}, \bibinfo {author}
  {\bibfnamefont {E.}~\bibnamefont {Arimondo}}, \bibinfo {author}
  {\bibfnamefont {D.}~\bibnamefont {Ciampini}}, \bibinfo {author}
  {\bibfnamefont {R.}~\bibnamefont {Fazio}}, \bibinfo {author} {\bibfnamefont
  {V.}~\bibnamefont {Giovannetti}}, \bibinfo {author} {\bibfnamefont
  {R.}~\bibnamefont {Mannella}},\ and\ \bibinfo {author} {\bibfnamefont
  {O.}~\bibnamefont {Morsch}},\ }\bibfield  {title} {\bibinfo {title}
  {High-fidelity quantum driving},\ }\href@noop {} {\bibfield  {journal}
  {\bibinfo  {journal} {Nature Physics}\ }\textbf {\bibinfo {volume} {8}},\
  \bibinfo {pages} {147} (\bibinfo {year} {2012})}\BibitemShut {NoStop}%
\bibitem [{\citenamefont {Lidar}\ and\ \citenamefont
  {Brun}(2013)}]{lidar2013quantum}%
  \BibitemOpen
  \bibfield  {author} {\bibinfo {author} {\bibfnamefont {D.~A.}\ \bibnamefont
  {Lidar}}\ and\ \bibinfo {author} {\bibfnamefont {T.~A.}\ \bibnamefont
  {Brun}},\ }\href@noop {} {\emph {\bibinfo {title} {Quantum error
  correction}}}\ (\bibinfo  {publisher} {Cambridge university press},\ \bibinfo
  {year} {2013})\BibitemShut {NoStop}%
\bibitem [{\citenamefont {Chang}\ \emph {et~al.}(2014)\citenamefont {Chang},
  \citenamefont {Xing}, \citenamefont {Zhang}, \citenamefont {Liu},
  \citenamefont {Jiang}, \citenamefont {Li}, \citenamefont {Gu}, \citenamefont
  {Long},\ and\ \citenamefont {Pan}}]{chang2014band}%
  \BibitemOpen
  \bibfield  {author} {\bibinfo {author} {\bibfnamefont {Y.-C.}\ \bibnamefont
  {Chang}}, \bibinfo {author} {\bibfnamefont {J.}~\bibnamefont {Xing}},
  \bibinfo {author} {\bibfnamefont {F.-H.}\ \bibnamefont {Zhang}}, \bibinfo
  {author} {\bibfnamefont {G.-Q.}\ \bibnamefont {Liu}}, \bibinfo {author}
  {\bibfnamefont {Q.-Q.}\ \bibnamefont {Jiang}}, \bibinfo {author}
  {\bibfnamefont {W.-X.}\ \bibnamefont {Li}}, \bibinfo {author} {\bibfnamefont
  {C.-Z.}\ \bibnamefont {Gu}}, \bibinfo {author} {\bibfnamefont {G.-L.}\
  \bibnamefont {Long}},\ and\ \bibinfo {author} {\bibfnamefont {X.-Y.}\
  \bibnamefont {Pan}},\ }\bibfield  {title} {\bibinfo {title} {Band-selective
  shaped pulse for high fidelity quantum control in diamond},\ }\href@noop {}
  {\bibfield  {journal} {\bibinfo  {journal} {Applied Physics Letters}\
  }\textbf {\bibinfo {volume} {104}},\ \bibinfo {pages} {262403} (\bibinfo
  {year} {2014})}\BibitemShut {NoStop}%
\bibitem [{\citenamefont {Spiteri}\ \emph {et~al.}(2018)\citenamefont
  {Spiteri}, \citenamefont {Schmidt}, \citenamefont {Ghosh}, \citenamefont
  {Zahedinejad},\ and\ \citenamefont {Sanders}}]{spiteri2018quantum}%
  \BibitemOpen
  \bibfield  {author} {\bibinfo {author} {\bibfnamefont {R.~J.}\ \bibnamefont
  {Spiteri}}, \bibinfo {author} {\bibfnamefont {M.}~\bibnamefont {Schmidt}},
  \bibinfo {author} {\bibfnamefont {J.}~\bibnamefont {Ghosh}}, \bibinfo
  {author} {\bibfnamefont {E.}~\bibnamefont {Zahedinejad}},\ and\ \bibinfo
  {author} {\bibfnamefont {B.~C.}\ \bibnamefont {Sanders}},\ }\bibfield
  {title} {\bibinfo {title} {Quantum control for high-fidelity multi-qubit
  gates},\ }\href@noop {} {\bibfield  {journal} {\bibinfo  {journal} {New
  Journal of Physics}\ }\textbf {\bibinfo {volume} {20}},\ \bibinfo {pages}
  {113009} (\bibinfo {year} {2018})}\BibitemShut {NoStop}%
\bibitem [{\citenamefont {Levy}\ \emph {et~al.}(2018)\citenamefont {Levy},
  \citenamefont {Kiely}, \citenamefont {Muga}, \citenamefont {Kosloff},\ and\
  \citenamefont {Torrontegui}}]{levy2018noise}%
  \BibitemOpen
  \bibfield  {author} {\bibinfo {author} {\bibfnamefont {A.}~\bibnamefont
  {Levy}}, \bibinfo {author} {\bibfnamefont {A.}~\bibnamefont {Kiely}},
  \bibinfo {author} {\bibfnamefont {J.~G.}\ \bibnamefont {Muga}}, \bibinfo
  {author} {\bibfnamefont {R.}~\bibnamefont {Kosloff}},\ and\ \bibinfo {author}
  {\bibfnamefont {E.}~\bibnamefont {Torrontegui}},\ }\bibfield  {title}
  {\bibinfo {title} {Noise resistant quantum control using dynamical
  invariants},\ }\href@noop {} {\bibfield  {journal} {\bibinfo  {journal} {New
  Journal of Physics}\ }\textbf {\bibinfo {volume} {20}},\ \bibinfo {pages}
  {025006} (\bibinfo {year} {2018})}\BibitemShut {NoStop}%
\bibitem [{\citenamefont {M\"ott\"onen}\ \emph {et~al.}(2006)\citenamefont
  {M\"ott\"onen}, \citenamefont {de~Sousa}, \citenamefont {Zhang},\ and\
  \citenamefont {Whaley}}]{PhysRevA.73.022332}%
  \BibitemOpen
  \bibfield  {author} {\bibinfo {author} {\bibfnamefont {M.}~\bibnamefont
  {M\"ott\"onen}}, \bibinfo {author} {\bibfnamefont {R.}~\bibnamefont
  {de~Sousa}}, \bibinfo {author} {\bibfnamefont {J.}~\bibnamefont {Zhang}},\
  and\ \bibinfo {author} {\bibfnamefont {K.~B.}\ \bibnamefont {Whaley}},\
  }\bibfield  {title} {\bibinfo {title} {High-fidelity one-qubit operations
  under random telegraph noise},\ }\href
  {https://doi.org/10.1103/PhysRevA.73.022332} {\bibfield  {journal} {\bibinfo
  {journal} {Phys. Rev. A}\ }\textbf {\bibinfo {volume} {73}},\ \bibinfo
  {pages} {022332} (\bibinfo {year} {2006})}\BibitemShut {NoStop}%
\bibitem [{\citenamefont {Daems}\ \emph {et~al.}(2013)\citenamefont {Daems},
  \citenamefont {Ruschhaupt}, \citenamefont {Sugny},\ and\ \citenamefont
  {Gu\'erin}}]{PhysRevLett.111.050404}%
  \BibitemOpen
  \bibfield  {author} {\bibinfo {author} {\bibfnamefont {D.}~\bibnamefont
  {Daems}}, \bibinfo {author} {\bibfnamefont {A.}~\bibnamefont {Ruschhaupt}},
  \bibinfo {author} {\bibfnamefont {D.}~\bibnamefont {Sugny}},\ and\ \bibinfo
  {author} {\bibfnamefont {S.}~\bibnamefont {Gu\'erin}},\ }\bibfield  {title}
  {\bibinfo {title} {Robust quantum control by a single-shot shaped pulse},\
  }\href {https://doi.org/10.1103/PhysRevLett.111.050404} {\bibfield  {journal}
  {\bibinfo  {journal} {Phys. Rev. Lett.}\ }\textbf {\bibinfo {volume} {111}},\
  \bibinfo {pages} {050404} (\bibinfo {year} {2013})}\BibitemShut {NoStop}%
\bibitem [{\citenamefont {Deng}\ \emph {et~al.}(2016)\citenamefont {Deng},
  \citenamefont {Shen}, \citenamefont {Ashhab},\ and\ \citenamefont
  {Lupascu}}]{PhysRevA.94.032323}%
  \BibitemOpen
  \bibfield  {author} {\bibinfo {author} {\bibfnamefont {C.}~\bibnamefont
  {Deng}}, \bibinfo {author} {\bibfnamefont {F.}~\bibnamefont {Shen}}, \bibinfo
  {author} {\bibfnamefont {S.}~\bibnamefont {Ashhab}},\ and\ \bibinfo {author}
  {\bibfnamefont {A.}~\bibnamefont {Lupascu}},\ }\bibfield  {title} {\bibinfo
  {title} {Dynamics of a two-level system under strong driving: Quantum-gate
  optimization based on floquet theory},\ }\href
  {https://doi.org/10.1103/PhysRevA.94.032323} {\bibfield  {journal} {\bibinfo
  {journal} {Phys. Rev. A}\ }\textbf {\bibinfo {volume} {94}},\ \bibinfo
  {pages} {032323} (\bibinfo {year} {2016})}\BibitemShut {NoStop}%
\bibitem [{\citenamefont {Yang}\ \emph {et~al.}(2019)\citenamefont {Yang},
  \citenamefont {Coppersmith},\ and\ \citenamefont
  {Friesen}}]{yang2019achieving}%
  \BibitemOpen
  \bibfield  {author} {\bibinfo {author} {\bibfnamefont {Y.-C.}\ \bibnamefont
  {Yang}}, \bibinfo {author} {\bibfnamefont {S.}~\bibnamefont {Coppersmith}},\
  and\ \bibinfo {author} {\bibfnamefont {M.}~\bibnamefont {Friesen}},\
  }\bibfield  {title} {\bibinfo {title} {Achieving high-fidelity single-qubit
  gates in a strongly driven charge qubit with 1/f charge noise},\ }\href@noop
  {} {\bibfield  {journal} {\bibinfo  {journal} {npj Quantum Information}\
  }\textbf {\bibinfo {volume} {5}},\ \bibinfo {pages} {1} (\bibinfo {year}
  {2019})}\BibitemShut {NoStop}%
\bibitem [{\citenamefont {Gershenfeld}\ and\ \citenamefont
  {Chuang}(1997)}]{gershenfeld1997bulk}%
  \BibitemOpen
  \bibfield  {author} {\bibinfo {author} {\bibfnamefont {N.~A.}\ \bibnamefont
  {Gershenfeld}}\ and\ \bibinfo {author} {\bibfnamefont {I.~L.}\ \bibnamefont
  {Chuang}},\ }\bibfield  {title} {\bibinfo {title} {Bulk spin-resonance
  quantum computation},\ }\href@noop {} {\bibfield  {journal} {\bibinfo
  {journal} {science}\ }\textbf {\bibinfo {volume} {275}},\ \bibinfo {pages}
  {350} (\bibinfo {year} {1997})}\BibitemShut {NoStop}%
\bibitem [{\citenamefont {Cory}\ \emph {et~al.}(1997)\citenamefont {Cory},
  \citenamefont {Fahmy},\ and\ \citenamefont {Havel}}]{cory1997ensemble}%
  \BibitemOpen
  \bibfield  {author} {\bibinfo {author} {\bibfnamefont {D.~G.}\ \bibnamefont
  {Cory}}, \bibinfo {author} {\bibfnamefont {A.~F.}\ \bibnamefont {Fahmy}},\
  and\ \bibinfo {author} {\bibfnamefont {T.~F.}\ \bibnamefont {Havel}},\
  }\bibfield  {title} {\bibinfo {title} {Ensemble quantum computing by nmr
  spectroscopy},\ }\href@noop {} {\bibfield  {journal} {\bibinfo  {journal}
  {Proceedings of the National Academy of Sciences}\ }\textbf {\bibinfo
  {volume} {94}},\ \bibinfo {pages} {1634} (\bibinfo {year}
  {1997})}\BibitemShut {NoStop}%
\bibitem [{\citenamefont {Vandersypen}\ \emph {et~al.}(2001)\citenamefont
  {Vandersypen}, \citenamefont {Steffen}, \citenamefont {Breyta}, \citenamefont
  {Yannoni}, \citenamefont {Sherwood},\ and\ \citenamefont
  {Chuang}}]{vandersypen2001experimental}%
  \BibitemOpen
  \bibfield  {author} {\bibinfo {author} {\bibfnamefont {L.~M.}\ \bibnamefont
  {Vandersypen}}, \bibinfo {author} {\bibfnamefont {M.}~\bibnamefont
  {Steffen}}, \bibinfo {author} {\bibfnamefont {G.}~\bibnamefont {Breyta}},
  \bibinfo {author} {\bibfnamefont {C.~S.}\ \bibnamefont {Yannoni}}, \bibinfo
  {author} {\bibfnamefont {M.~H.}\ \bibnamefont {Sherwood}},\ and\ \bibinfo
  {author} {\bibfnamefont {I.~L.}\ \bibnamefont {Chuang}},\ }\bibfield  {title}
  {\bibinfo {title} {Experimental realization of shor's quantum factoring
  algorithm using nuclear magnetic resonance},\ }\href@noop {} {\bibfield
  {journal} {\bibinfo  {journal} {Nature}\ }\textbf {\bibinfo {volume} {414}},\
  \bibinfo {pages} {883} (\bibinfo {year} {2001})}\BibitemShut {NoStop}%
\bibitem [{\citenamefont {Kondo}(2009)}]{kondo2009liquid}%
  \BibitemOpen
  \bibfield  {author} {\bibinfo {author} {\bibfnamefont {Y.}~\bibnamefont
  {Kondo}},\ }\bibfield  {title} {\bibinfo {title} {Liquid-state nmr quantum
  computer: Working principle and some examples},\ }in\ \href@noop {} {\emph
  {\bibinfo {booktitle} {Molecular Realizations of Quantum Computing 2007}}}\
  (\bibinfo  {publisher} {World Scientific},\ \bibinfo {year} {2009})\ pp.\
  \bibinfo {pages} {1--52}\BibitemShut {NoStop}%
\bibitem [{\citenamefont {Jones}(2011)}]{JONES201191}%
  \BibitemOpen
  \bibfield  {author} {\bibinfo {author} {\bibfnamefont {J.~A.}\ \bibnamefont
  {Jones}},\ }\bibfield  {title} {\bibinfo {title} {Quantum computing with
  nmr},\ }\href {https://doi.org/https://doi.org/10.1016/j.pnmrs.2010.11.001}
  {\bibfield  {journal} {\bibinfo  {journal} {Progress in Nuclear Magnetic
  Resonance Spectroscopy}\ }\textbf {\bibinfo {volume} {59}},\ \bibinfo {pages}
  {91} (\bibinfo {year} {2011})}\BibitemShut {NoStop}%
\bibitem [{\citenamefont {Counsell}\ \emph {et~al.}(1985)\citenamefont
  {Counsell}, \citenamefont {Levitt},\ and\ \citenamefont
  {Ernst}}]{counsell1985analytical}%
  \BibitemOpen
  \bibfield  {author} {\bibinfo {author} {\bibfnamefont {C.}~\bibnamefont
  {Counsell}}, \bibinfo {author} {\bibfnamefont {M.}~\bibnamefont {Levitt}},\
  and\ \bibinfo {author} {\bibfnamefont {R.}~\bibnamefont {Ernst}},\ }\bibfield
   {title} {\bibinfo {title} {Analytical theory of composite pulses},\
  }\href@noop {} {\bibfield  {journal} {\bibinfo  {journal} {Journal of
  Magnetic Resonance (1969)}\ }\textbf {\bibinfo {volume} {63}},\ \bibinfo
  {pages} {133} (\bibinfo {year} {1985})}\BibitemShut {NoStop}%
\bibitem [{\citenamefont {Levitt}(1986)}]{levitt1986composite}%
  \BibitemOpen
  \bibfield  {author} {\bibinfo {author} {\bibfnamefont {M.~H.}\ \bibnamefont
  {Levitt}},\ }\bibfield  {title} {\bibinfo {title} {Composite pulses},\
  }\href@noop {} {\bibfield  {journal} {\bibinfo  {journal} {Progress in
  Nuclear Magnetic Resonance Spectroscopy}\ }\textbf {\bibinfo {volume} {18}},\
  \bibinfo {pages} {61} (\bibinfo {year} {1986})}\BibitemShut {NoStop}%
\bibitem [{\citenamefont {Claridge}(2016)}]{claridge2016high}%
  \BibitemOpen
  \bibfield  {author} {\bibinfo {author} {\bibfnamefont {T.~D.}\ \bibnamefont
  {Claridge}},\ }\href@noop {} {\emph {\bibinfo {title} {High-resolution NMR
  techniques in organic chemistry}}},\ Vol.~\bibinfo {volume} {27}\ (\bibinfo
  {publisher} {Elsevier},\ \bibinfo {year} {2016})\BibitemShut {NoStop}%
\bibitem [{\citenamefont {Brown}\ \emph {et~al.}(2004)\citenamefont {Brown},
  \citenamefont {Harrow},\ and\ \citenamefont {Chuang}}]{brown2004arbitrarily}%
  \BibitemOpen
  \bibfield  {author} {\bibinfo {author} {\bibfnamefont {K.~R.}\ \bibnamefont
  {Brown}}, \bibinfo {author} {\bibfnamefont {A.~W.}\ \bibnamefont {Harrow}},\
  and\ \bibinfo {author} {\bibfnamefont {I.~L.}\ \bibnamefont {Chuang}},\
  }\bibfield  {title} {\bibinfo {title} {Arbitrarily accurate composite pulse
  sequences},\ }\href@noop {} {\bibfield  {journal} {\bibinfo  {journal}
  {Physical Review A}\ }\textbf {\bibinfo {volume} {70}},\ \bibinfo {pages}
  {052318} (\bibinfo {year} {2004})}\BibitemShut {NoStop}%
\bibitem [{\citenamefont {Wimperis}(1994)}]{wimperis1994broadband}%
  \BibitemOpen
  \bibfield  {author} {\bibinfo {author} {\bibfnamefont {S.}~\bibnamefont
  {Wimperis}},\ }\bibfield  {title} {\bibinfo {title} {Broadband, narrowband,
  and passband composite pulses for use in advanced nmr experiments},\
  }\href@noop {} {\bibfield  {journal} {\bibinfo  {journal} {Journal of
  Magnetic Resonance, Series A}\ }\textbf {\bibinfo {volume} {109}},\ \bibinfo
  {pages} {221} (\bibinfo {year} {1994})}\BibitemShut {NoStop}%
\bibitem [{\citenamefont {Cummins}\ \emph {et~al.}(2003)\citenamefont
  {Cummins}, \citenamefont {Llewellyn},\ and\ \citenamefont
  {Jones}}]{cummins2003tackling}%
  \BibitemOpen
  \bibfield  {author} {\bibinfo {author} {\bibfnamefont {H.~K.}\ \bibnamefont
  {Cummins}}, \bibinfo {author} {\bibfnamefont {G.}~\bibnamefont {Llewellyn}},\
  and\ \bibinfo {author} {\bibfnamefont {J.~A.}\ \bibnamefont {Jones}},\
  }\bibfield  {title} {\bibinfo {title} {Tackling systematic errors in quantum
  logic gates with composite rotations},\ }\href@noop {} {\bibfield  {journal}
  {\bibinfo  {journal} {Physical Review A}\ }\textbf {\bibinfo {volume} {67}},\
  \bibinfo {pages} {042308} (\bibinfo {year} {2003})}\BibitemShut {NoStop}%
\bibitem [{\citenamefont {Torosov}\ and\ \citenamefont
  {Vitanov}(2011)}]{PhysRevA.83.053420}%
  \BibitemOpen
  \bibfield  {author} {\bibinfo {author} {\bibfnamefont {B.~T.}\ \bibnamefont
  {Torosov}}\ and\ \bibinfo {author} {\bibfnamefont {N.~V.}\ \bibnamefont
  {Vitanov}},\ }\bibfield  {title} {\bibinfo {title} {Smooth composite pulses
  for high-fidelity quantum information processing},\ }\href
  {https://doi.org/10.1103/PhysRevA.83.053420} {\bibfield  {journal} {\bibinfo
  {journal} {Phys. Rev. A}\ }\textbf {\bibinfo {volume} {83}},\ \bibinfo
  {pages} {053420} (\bibinfo {year} {2011})}\BibitemShut {NoStop}%
\bibitem [{\citenamefont {Torosov}\ \emph {et~al.}(2011)\citenamefont
  {Torosov}, \citenamefont {Gu\'erin},\ and\ \citenamefont
  {Vitanov}}]{PhysRevLett.106.233001}%
  \BibitemOpen
  \bibfield  {author} {\bibinfo {author} {\bibfnamefont {B.~T.}\ \bibnamefont
  {Torosov}}, \bibinfo {author} {\bibfnamefont {S.}~\bibnamefont {Gu\'erin}},\
  and\ \bibinfo {author} {\bibfnamefont {N.~V.}\ \bibnamefont {Vitanov}},\
  }\bibfield  {title} {\bibinfo {title} {High-fidelity adiabatic passage by
  composite sequences of chirped pulses},\ }\href
  {https://doi.org/10.1103/PhysRevLett.106.233001} {\bibfield  {journal}
  {\bibinfo  {journal} {Phys. Rev. Lett.}\ }\textbf {\bibinfo {volume} {106}},\
  \bibinfo {pages} {233001} (\bibinfo {year} {2011})}\BibitemShut {NoStop}%
\bibitem [{\citenamefont {Jones}(2013)}]{jones2013designing}%
  \BibitemOpen
  \bibfield  {author} {\bibinfo {author} {\bibfnamefont {J.~A.}\ \bibnamefont
  {Jones}},\ }\bibfield  {title} {\bibinfo {title} {Designing short robust not
  gates for quantum computation},\ }\href@noop {} {\bibfield  {journal}
  {\bibinfo  {journal} {Physical Review A}\ }\textbf {\bibinfo {volume} {87}},\
  \bibinfo {pages} {052317} (\bibinfo {year} {2013})}\BibitemShut {NoStop}%
\bibitem [{\citenamefont {Genov}\ \emph {et~al.}(2014)\citenamefont {Genov},
  \citenamefont {Schraft}, \citenamefont {Halfmann},\ and\ \citenamefont
  {Vitanov}}]{PhysRevLett.113.043001}%
  \BibitemOpen
  \bibfield  {author} {\bibinfo {author} {\bibfnamefont {G.~T.}\ \bibnamefont
  {Genov}}, \bibinfo {author} {\bibfnamefont {D.}~\bibnamefont {Schraft}},
  \bibinfo {author} {\bibfnamefont {T.}~\bibnamefont {Halfmann}},\ and\
  \bibinfo {author} {\bibfnamefont {N.~V.}\ \bibnamefont {Vitanov}},\
  }\bibfield  {title} {\bibinfo {title} {Correction of arbitrary field errors
  in population inversion of quantum systems by universal composite pulses},\
  }\href {https://doi.org/10.1103/PhysRevLett.113.043001} {\bibfield  {journal}
  {\bibinfo  {journal} {Phys. Rev. Lett.}\ }\textbf {\bibinfo {volume} {113}},\
  \bibinfo {pages} {043001} (\bibinfo {year} {2014})}\BibitemShut {NoStop}%
\bibitem [{\citenamefont {Kyoseva}\ \emph {et~al.}(2019)\citenamefont
  {Kyoseva}, \citenamefont {Greener},\ and\ \citenamefont
  {Suchowski}}]{PhysRevA.100.032333}%
  \BibitemOpen
  \bibfield  {author} {\bibinfo {author} {\bibfnamefont {E.}~\bibnamefont
  {Kyoseva}}, \bibinfo {author} {\bibfnamefont {H.}~\bibnamefont {Greener}},\
  and\ \bibinfo {author} {\bibfnamefont {H.}~\bibnamefont {Suchowski}},\
  }\bibfield  {title} {\bibinfo {title} {Detuning-modulated composite pulses
  for high-fidelity robust quantum control},\ }\href
  {https://doi.org/10.1103/PhysRevA.100.032333} {\bibfield  {journal} {\bibinfo
   {journal} {Phys. Rev. A}\ }\textbf {\bibinfo {volume} {100}},\ \bibinfo
  {pages} {032333} (\bibinfo {year} {2019})}\BibitemShut {NoStop}%
\bibitem [{\citenamefont {Cummins}\ and\ \citenamefont
  {Jones}(2000)}]{cummins2000use}%
  \BibitemOpen
  \bibfield  {author} {\bibinfo {author} {\bibfnamefont {H.}~\bibnamefont
  {Cummins}}\ and\ \bibinfo {author} {\bibfnamefont {J.}~\bibnamefont
  {Jones}},\ }\bibfield  {title} {\bibinfo {title} {Use of composite rotations
  to correct systematic errors in nmr quantum computation},\ }\href@noop {}
  {\bibfield  {journal} {\bibinfo  {journal} {New Journal of Physics}\ }\textbf
  {\bibinfo {volume} {2}},\ \bibinfo {pages} {6} (\bibinfo {year}
  {2000})}\BibitemShut {NoStop}%
\bibitem [{\citenamefont {M{\"o}tt{\"o}nen}\ \emph {et~al.}(2006)\citenamefont
  {M{\"o}tt{\"o}nen}, \citenamefont {de~Sousa}, \citenamefont {Zhang},\ and\
  \citenamefont {Whaley}}]{mottonen2006high}%
  \BibitemOpen
  \bibfield  {author} {\bibinfo {author} {\bibfnamefont {M.}~\bibnamefont
  {M{\"o}tt{\"o}nen}}, \bibinfo {author} {\bibfnamefont {R.}~\bibnamefont
  {de~Sousa}}, \bibinfo {author} {\bibfnamefont {J.}~\bibnamefont {Zhang}},\
  and\ \bibinfo {author} {\bibfnamefont {K.~B.}\ \bibnamefont {Whaley}},\
  }\bibfield  {title} {\bibinfo {title} {High-fidelity one-qubit operations
  under random telegraph noise},\ }\href@noop {} {\bibfield  {journal}
  {\bibinfo  {journal} {Physical Review A}\ }\textbf {\bibinfo {volume} {73}},\
  \bibinfo {pages} {022332} (\bibinfo {year} {2006})}\BibitemShut {NoStop}%
\bibitem [{\citenamefont {Said}\ and\ \citenamefont
  {Twamley}(2009)}]{said2009robust}%
  \BibitemOpen
  \bibfield  {author} {\bibinfo {author} {\bibfnamefont {R.}~\bibnamefont
  {Said}}\ and\ \bibinfo {author} {\bibfnamefont {J.}~\bibnamefont {Twamley}},\
  }\bibfield  {title} {\bibinfo {title} {Robust control of entanglement in a
  nitrogen-vacancy center coupled to a c 13 nuclear spin in diamond},\
  }\href@noop {} {\bibfield  {journal} {\bibinfo  {journal} {Physical Review
  A}\ }\textbf {\bibinfo {volume} {80}},\ \bibinfo {pages} {032303} (\bibinfo
  {year} {2009})}\BibitemShut {NoStop}%
\bibitem [{\citenamefont {Timoney}\ \emph {et~al.}(2008)\citenamefont
  {Timoney}, \citenamefont {Elman}, \citenamefont {Glaser}, \citenamefont
  {Weiss}, \citenamefont {Johanning}, \citenamefont {Neuhauser},\ and\
  \citenamefont {Wunderlich}}]{timoney2008error}%
  \BibitemOpen
  \bibfield  {author} {\bibinfo {author} {\bibfnamefont {N.}~\bibnamefont
  {Timoney}}, \bibinfo {author} {\bibfnamefont {V.}~\bibnamefont {Elman}},
  \bibinfo {author} {\bibfnamefont {S.}~\bibnamefont {Glaser}}, \bibinfo
  {author} {\bibfnamefont {C.}~\bibnamefont {Weiss}}, \bibinfo {author}
  {\bibfnamefont {M.}~\bibnamefont {Johanning}}, \bibinfo {author}
  {\bibfnamefont {W.}~\bibnamefont {Neuhauser}},\ and\ \bibinfo {author}
  {\bibfnamefont {C.}~\bibnamefont {Wunderlich}},\ }\bibfield  {title}
  {\bibinfo {title} {Error-resistant single-qubit gates with trapped ions},\
  }\href@noop {} {\bibfield  {journal} {\bibinfo  {journal} {Physical Review
  A}\ }\textbf {\bibinfo {volume} {77}},\ \bibinfo {pages} {052334} (\bibinfo
  {year} {2008})}\BibitemShut {NoStop}%
\bibitem [{\citenamefont {Bando}\ \emph {et~al.}(2012)\citenamefont {Bando},
  \citenamefont {Ichikawa}, \citenamefont {Kondo},\ and\ \citenamefont
  {Nakahara}}]{bando2012concatenated}%
  \BibitemOpen
  \bibfield  {author} {\bibinfo {author} {\bibfnamefont {M.}~\bibnamefont
  {Bando}}, \bibinfo {author} {\bibfnamefont {T.}~\bibnamefont {Ichikawa}},
  \bibinfo {author} {\bibfnamefont {Y.}~\bibnamefont {Kondo}},\ and\ \bibinfo
  {author} {\bibfnamefont {M.}~\bibnamefont {Nakahara}},\ }\bibfield  {title}
  {\bibinfo {title} {Concatenated composite pulses compensating simultaneous
  systematic errors},\ }\href@noop {} {\bibfield  {journal} {\bibinfo
  {journal} {Journal of the Physical Society of Japan}\ }\textbf {\bibinfo
  {volume} {82}},\ \bibinfo {pages} {014004} (\bibinfo {year}
  {2012})}\BibitemShut {NoStop}%
\end{thebibliography}%
	\appendix
	
	\section{Time-symmetric $k=3$ ORE-robust CPs}
	\label{appendix1}
	
	\subsection{time symmetry and ORE robustness}
	
	First, we show the relationship between the time symmetry of $k=3$ CPs and the ORE robustness.
	Consider a general form of $k=3$ CPs:
	\begin{equation}
		\bar{U}_{3}\bar{U}_{2}\bar{U}_{1}:=U(\theta_{3},\phi_{3})U(\theta_{2},\phi_{2})U(\theta_{1},\phi_{1}),
	\end{equation}
	where $\bar{U}_{i}:=U(\theta_{i},\phi_{i})$.
	The condition that the above CP is robust against the ORE is
	\begin{equation}
		\sin(\theta_{3}/2)\sigma_{z}\bar{U}_{2}\bar{U}_{1}+\sin(\theta_{2}/2)\bar{U}_{3}\sigma_{z}\bar{U}_{1}+\sin(\theta_{1}/2)\bar{U}_{3}\bar{U}_{2}\sigma_{z}=0.
	\end{equation}
	By operating $\bar{U}^{\dagger}_{1}$ from r.h.s. and $\bar{U}^{\dagger}_{3}$ from l.h.s.,
	we obtain
	\begin{equation}
		\sin(\theta_{2}/2)\sigma_{z}+\sin(\theta_{3}/2)\bar{U}^{\dagger}_{3}\sigma_{z}\bar{U}_{2}+\sin(\theta_{1}/2)\bar{U}_{2}\sigma_{z}\bar{U}^{\dagger}_{1}=0.
	\end{equation}
	This implies
	\begin{equation}
		\sin(\theta_{2}/2){\mathbb I}+\sin(\theta_{3}/2)\bar{U}_{3}\bar{U}_{2}+\sin(\theta_{1}/2)\bar{U}^{\dagger}_{2}\bar{U}^{\dagger}_{1}=0,
		\label{eq:condition_for_k=3}
	\end{equation}
	where we use $\sigma_{z}\bar{U}_{i}=\bar{U}^{\dagger}_{i}\sigma_{z}$ ($\vec{n}_{i}$ is orthogonal to $\vec{z}$ for any $i$).
	To further deform Eq. (\ref{eq:condition_for_k=3}), we evaluate the following equation:
	\begin{equation}
		A+B=\alpha{\mathbb I},
		\label{eq:unity}
	\end{equation}
	where $\alpha$ is a nonzero number, and $A$ and $B$ are matrices.
	By simple calculations, we can prove that $A$ and $B$ commute; that is, $[A,B]=0$:
	\begin{equation}
		A+B=\alpha{\mathbb I}~~\Longrightarrow~~B=\alpha {\mathbb I}-A~~\Longrightarrow~~[A,B]=0.
	\end{equation}
	Applying this fact to Eq.~(\ref{eq:condition_for_k=3}), we obtain
	\begin{equation}
		[\sin(\theta_{3}/2)\bar{U}_{3}\bar{U}_{2},\sin(\theta_{1}/2)\bar{U}^{\dagger}_{2}\bar{U}^{\dagger}_{1}]=0,
	\end{equation}
	which is a {\it necessary} condition for the ORE robustness.
	Note that $\sin(\theta_{1}/2)$ ($\sin(\theta_{3}/2)$) is zero only when the first (third) operation is a $2 n \pi$-rotation.
	Because such an operation cannot contribute to the (first-order) ORE robustness (see the ORE robustness condition (\ref{eq:condition_for_k=3})), we assume that $\sin(\theta_{1}/2)$ and $\sin(\theta_{3}/2)$ are nonzero.
	Then straightforward calculations show that the above equation is equivalent to the following:
	\begin{equation}
		\bar{U}_{3}\bar{U}_{2}\bar{U}_{1}=\bar{U}_{1}\bar{U}_{2}\bar{U}_{3}.
		\label{eq:presymmet}
	\end{equation}
	The simplest case for satisfying Eq. (\ref{eq:presymmet}) is $\bar{U}_{3}=\pm \bar{U}_{1}$, or equivalently,
	\begin{equation}
		\theta_{3}=\theta_{1}+2 n \pi,~~\phi_{1}=\phi_{3},~~n \in {\mathbb Z},~~\theta_{1}, \theta_{3}>0.
		\label{eq:symmet}
	\end{equation}
	The first and third operations are symmetric the same modulo $2\pi$ of $\theta$.
	Therefore, we consider the symmetric ORE-robust CPs in the main text.
	
	\subsection{time symmetry and target operation direction}
	
	Here, we show that the condition (\ref{eq:presymmet}) also restricts the target operation to the form of $U(\theta,\phi)$.
	The direction $\vec{n}$ of the target operation is restricted into the $xy$ plane.
	When each elementary operation has the form of $U(\theta_{i},\phi_{i})$, some calculations show that $\bar{U}_{1}\bar{U}_{2}\bar{U}_{3}$ and $\bar{U}_{3}\bar{U}_{2}\bar{U}_{1}$ can be written as
	\begin{equation}
		\bar{U}_{1}\bar{U}_{2}\bar{U}_{3}=
		\begin{pmatrix}
			A+iB&C\\
			C^{*}&A-iB
		\end{pmatrix}
		,~~
		\bar{U}_{3}\bar{U}_{2}\bar{U}_{1}=
		\begin{pmatrix}
			A-iB&C\\
			C^{*}&A+iB
		\end{pmatrix}
		,
	\end{equation}
	with the same coefficients, $A,B \in {\mathbb R}$, and $C\in{\mathbb C}$.
	The condition (\ref{eq:presymmet}) directly implies $B=0$.
	Comparing the form of the elementary operation (\ref{eq:elementary}),
	we find that the direction vector $\vec{n}$ of $\bar{U}_{1}\bar{U}_{2}\bar{U}_{3}(=\bar{U}_{3}\bar{U}_{2}\bar{U}_{1})$ is in the $xy$ plane;
	$B$ is the coefficient of $\sigma_{z}$.
	Accordingly, the symmetric condition $\bar{U}_{3}=\pm \bar{U}_{1}$ restricts the form of the target operation.

	\section{Structure of Eq. (\ref{eq:target2})}
	\label{appendix2}
	
	\subsection{radial equation}
	
	Because Eq. (\ref{eq:target2}) is a complex equation, it can be decomposed into radial and phase equations.
	Here, we demonstrate that the radial part is always satisfied when Eq. (\ref{eq:target1}) holds.
	The squared absolute value of lhs in Eq. (\ref{eq:target2}) is calculated as
	\begin{align}
		|2 s_{1}c_{1}c_{2}e^{-i (\phi_{2}-\phi_{1})}+c^{2}_{1}s_{2}-s^{2}_{1}s_{2}e^{-2i(\phi_{2}-\phi_{1})}|^{2}=&(2s_{1}c_{1}c_{2})^{2}+c^{4}_{1}s^{2}_{2}+s^{4}_{1}s^{2}_{2}+2c^{2}_{1}s^{2}_{1}s^{2}_{1}\nonumber\\
		&+4(c^{2}_{1}-s^{2}_{1})s_{1}c_{1}s_{2}c_{2}\alpha-4 s^{2}_{1}c^{2}_{1} s^{2}_{2}\alpha^{2}\nonumber\\
		=&1-c^{2}_{2}+4 s^{2}_{1}c^{2}_{1}c^{2}_{2}+4(c^{2}_{1}-s^{2}_{1})s_{1}c_{1}s_{2}c_{2}\alpha-4 s^{2}_{1}c^{2}_{1} s^{2}_{2}\alpha^{2}\nonumber\\
		=&1-\bigl( c_{2} (c^{2}_{2}-s^{2}_{2}-2 \alpha s_{1}c_{1}s_{2})\bigr)^{2}=1-c^{2}=s^{2}
	\end{align}
	where we use Eq. (\ref{eq:target1}) in the last line.
	The absolute value of lhs in Eq. (\ref{eq:target2}) is identical to that of rhs owing to Eq. (\ref{eq:target1}).
	Thus, we have shown that Eq.~(\ref{eq:target2}) is actually equivalent to one real equation when Eq.~(\ref{eq:target1}) is satisfied.
	The number of the original variables ${c_{1},c_{2},k,l}$ is four, while the number of the actual equations that must be solved is three.
	This is the reason why we can freely choose $c_{1}$.
	
	\subsection{phase equation}
	
	After solving Eqs. (\ref{eq:ORErobust}) and (\ref{eq:target1}), the variable $k=\phi_{2}-\phi$ is determined by the following equations equivalent to Eq. (\ref{eq:target2}):
	\begin{align}
		\cos k=&\frac{(-1)^{n}}{s}\bigl(2 s_{1}c_{1}c_{2}\cos l +c^{2}_{1}s_{2}-s^{2}_{1}s_{2}\cos(2l)\bigr)\nonumber\\
		\sin k=&\frac{(-1)^{n}}{s}\bigl(2 s_{1}c_{1}c_{2}\sin l-s^{2}_{1}s_{2}\sin(2l)\bigr),
	\end{align}
	where we use $l=\phi_{2}-\phi_{1}$.
	These equations always have a solution $k$ because of the radial identity.
	Note that Eqs. (\ref{eq:ORErobust}) and ($\ref{eq:target1}$) cannot determine $l$ itself:
	they just fix $\alpha =\cos l$.
	This means that the signature of $\sin l$ is arbitrary.
	For either choice of signature, we obtain a different $k$, but both choices are valid for constructing an ORE-robust CP.
	
	\section{Derivation of Equations (\ref{eq:solution1},\ref{eq:solution2},\ref{eq:bound})}
	\label{appendix3}
	
	Let us solve Eqs. (\ref{eq:ORErobust}) and (\ref{eq:target1}), and derive the solutions (\ref{eq:solution1}) and (\ref{eq:solution2}).
	For this purpose, it is convenient to introduce $c_{n}:=(-1)^{n}c$.
	Eqs. (\ref{eq:ORErobust}) and (\ref{eq:target1}) can be rewritten as
	\begin{align}
		&s_{2}+s_{1}(c_{2}c_{1}-\alpha s_{2}s_{1})=0,\label{eq:solv1}\\
		&c_{2}(c^{2}_{1}-s^{2}_{1})-2\alpha c_{1}s_{1}s_{2}= c_{n},
		\label{eq:solv2}
	\end{align}
	with $c_{n}$.
	These two equations contain three variables $(c_{1},c_{2},\alpha)$; hence, one parameter cannot be determined.
	We take $c_{1}$ as the free parameter without loss of generality.
	Our purpose is to represent $\alpha$ and $c_{2}$ as a function of $c$ and $c_{1}$.
	According to Eq. (\ref{eq:solv1}), $\alpha$ can be expressed as
	\begin{equation}
		\alpha=\frac{1}{s^{2}_{1}}+\frac{c_{1}c_{2}}{s_{1}s_{2}},
		\label{eq:alpha_tentative}
	\end{equation}
	where we assume $s_{1,2}\neq0$ because $s_{i}=0$ implies that $\theta_{i}=2 \pi n~(n\in {\mathbb Z})$, and such an operation cannot contribute to the first-order ORE robustness.
	Substituting this equation into Eq. (\ref{eq:solv2}), we obtain
	\begin{equation}
		c_{2}+\frac{c_{1}s_{2}}{s_{1}}=-c_{n}.
	\end{equation}
	As $s_{i}$'s are positive and written as $s_{i}=\sqrt{1-c^{2}_{i}}$, the above equation is rewritten as
	\begin{equation}
		c_{1}\sqrt{1-c^{2}_{2}}=-(c_{n}+c_{2})\sqrt{1-c^{2}_{1}}.
		\label{eq:c2equation}
	\end{equation}
	Taking the square of both sides, we obtain
	\begin{equation}
		c^{2}_{2}+2c_{n}(1-c^{2}_{1})c_{2}-c^{2}_{1}+(1-c^{2}_{1})c^{2}=0.
		\label{eq:c2sqequation}
	\end{equation}
	We solve this equation with respect to $c_{2}$ and obtain
	\begin{equation}
		c_{2}=-c_{n}(1-c^{2}_{1})\pm|c_{1}|\sqrt{1-s^{2}_{1}c^{2}}.
	\end{equation}
	Note that these are the solutions of the squared equation (\ref{eq:c2sqequation}).
	It is necessary to verify the sign that should be chosen for Eq. (\ref{eq:c2equation}).
	Substituting this solution into the original equation (\ref{eq:c2equation}), we obtain
	\begin{equation}
		c_{1}\sqrt{1-c^{2}_{2}}=-\Bigl(c_{n}c^{2}_{1}\pm|c_{1}|\sqrt{1-s^{2}_{1}c^{2}}\Bigr)\sqrt{1-c^{2}_{1}}.
	\end{equation}
	Note that the following relation is satisfied:
	\begin{equation}
	\sqrt{1-s^{2}_{1}c^{2}}=\sqrt{c^{2}c^{2}_{1}+(1-c^{2})}\geq|c_{1}c|=|c_{1}c_{n}|.
	\label{eq:inequality}
	\end{equation}
 	When $c_{1}$ is positive, the lhs is also positive.
 	Thus, to guarantee consistency, we must take the minus sign in rhs, regardless of the other parameters $c$ and $n$.
 	Similarly, when $c_{1}$ is negative, the sign must be positive.
 	Hence, $c_{2}$ is given as
 	\begin{equation}
 		c_{2}=-c_{n}(1-c^{2}_{1})-c_{1}\sqrt{1-s^{2}_{1}c^{2}}=-(-1)^{n}c(1-c^{2}_{1})-c_{1}\sqrt{1-s^{2}_{1}c^{2}}.
 		\label{eq:true_c2}
 	\end{equation}
 	The combination of the minus (plus) sign and absolute value $|c_{1}|$ with positive (negative) $c_{1}$ can always be written as $-c_{1}$. 
 	
	Calculating $s_{2}$ through the relation $s_{2}=\sqrt{1-c^{2}_{2}}$, we obtain
	\begin{align}
		s_{2}&=\sqrt{1-\Bigl(-c_{n}(1-c^{2}_{1})-c_{1}\sqrt{1-s^{2}_{1}c^{2}} \Bigr)^{2}}\nonumber\\
		&=|s_{1}|\sqrt{1-s^{2}_{1}c^{2}-2c_{n}c_{1}\sqrt{1-s^{2}_{1}c^{2}}+c^{2}_{1}c^{2}}\nonumber\\
		&=|s_{1}|\sqrt{\Bigl(\sqrt{1-s^{2}_{1}c^{2}}-c_{n}c_{1} \Bigr)^{2}}.
	\end{align}
	Using the positivity of $s_{1}$ and Eq. (\ref{eq:inequality}), we obtain
	\begin{equation}
		s_{2}=s_{1}\Bigl(\sqrt{1-s^{2}_{1}c^{2}}-c_{n}c_{1} \Bigr)=s_{1}\Bigl(\sqrt{1-s^{2}_{1}c^{2}}-(-1)^{n}c c_{1} \Bigr).
		\label{eq:true_s2}
	\end{equation}
	We obtain the explicit form of $\alpha$ by substituting Eqs. (\ref{eq:true_c2}) and (\ref{eq:true_s2}) into Eq. (\ref{eq:alpha_tentative}):
	\begin{equation}
		\alpha=1-\frac{\sqrt{1-s^{2}_{1}c^{2}}+(-1)^{n}c c_{1}}{2 s^{2}_{1}\bigl(\sqrt{1-s^{2}_{1}c^{2}}-(-1)^{n}c c_{1}\bigr)}.
		\label{true_alpha}
	\end{equation}

	Then, we explain the origin of the bound (\ref{eq:bound}).
	Note that $\alpha$ must satisfy $-1\leq \alpha \leq 1$ because it is originally the inner product of $\vec{n}_{1}$ and $\vec{n}_{2}$.
	The second term in Eq. (\ref{true_alpha}) is non-negative; hence, the condition $\alpha\leq 1$ is always satisfied.
	
	On the other hand, the condition $\alpha \geq -1$ provides a nontrivial constraint for allowed values of $c_{1}$ as a function of $c$.
	Here, we show how this constrains $c_{1}$.
	The condition of $\alpha \geq -1$ can be written as
	\begin{equation}
		(4 c^{2}_{1}-3)\sqrt{1-s^{2}_{1}c^{2}}\leq-(5-4c^{2}_{1})c_{n}c_{1}.
		\label{eq:ineq0}
	\end{equation}
	Note that $5-4 c^{2}_{1}$ is always positive and the sign of both sides varies depending on $c_{1}$ as follows:
	\begin{itemize}
		\item $-1 \leq c_{1} \leq -\frac{\sqrt{3}}{2}$ $\Longrightarrow$ lhs: non-negative, rhs: ${\rm sgn}(c_{n})$,
		\item $ -\frac{\sqrt{3}}{2} < c_{1} \leq 0$  ~~$\Longrightarrow$ lhs: negative, rhs: ${\rm sgn}(c_{n})$,
		\item $ 0 < c_{1} < \frac{\sqrt{3}}{2}$  ~~~~$\Longrightarrow$ lhs: negative, rhs: $-{\rm sgn}(c_{n})$,
		\item $ \frac{\sqrt{3}}{2} \leq c_{1} \leq 1$  ~~~~$\Longrightarrow$ lhs: non-negative, rhs: $-{\rm sgn}(c_{n})$.
		\label{eq:cases}
	\end{itemize}
	Note that $-1\leq c_{1}:=\cos(\theta^{(p)}_{1}/2)\leq 1$.
	We solve the inequality according to this division.
	
	Hereinafter we focus only on the case of $c_{n} \geq 0$ because the other case $c_{n}\leq 0$ is easily solved similarly.
	When $c_{n}\geq 0$, only the top three equations are valid because a positive value cannot be smaller than a negative value.
	
	\subsection{case of $-1 \leq c_{1} \leq -\frac{\sqrt{3}}{2}$}
	\label{subsec1}
	
	In this case, both sides of Eq. (\ref{eq:ineq0}) are positive.
	By squaring both sides, we obtain the following form of the inequality:
	\begin{equation}
		16 c^{4}_{1} -8(2+s^{2})c^{2}_{1}+9 s^{2} \leq 0.
	\end{equation}
	This inequality leads to
	\begin{align}
		&-g_{+}(s)\leq c_{1} \leq -g_{-}(s),~{\rm or}~g_{-}(s)\leq c_{1} \leq g_{+}(s),\nonumber\\
		&g_{-}(s)=\frac{\sqrt{2+s^{2}-\sqrt{s^{4}-5s^{2}+4}}}{2},~~g_{+}(s)=\frac{\sqrt{2+s^{2}+\sqrt{s^{4}-5s^{2}+4}}}{2}.
		\label{eq:ineq01}
	\end{align}
	For later convenience, we calculate the behaviour of $g_{-}(s)$ and $g_{+}(s)$.
	First, note that the numerator of $g_{-}(s)$ and $g_{+}(s)$ are always real, because it is easy to check that $|2+s^{2}|\geq \sqrt{s^{4}-5s^{2}+4}$.
	Hence, the inequality (\ref{eq:ineq01}) is always valid.
	The derivative of $\bigl(g_{-}(s)\bigr)^{2}$ with respect to $s^{2}$ is
	\begin{equation}
		\frac{d \bigl(g_{-}(s)\bigr)^{2}}{d (s^{2})}=\frac{2\sqrt{s^{4}-5s^{2}+4}-(2s^{2}-5)}{8 \sqrt{s^{4}-5s^{2}+4}}
	\end{equation}
	It is easily determined that the numerator is always positive.
	Thus, $\bigl(g_{-}(s)\bigr)^{2}$ is a monotonically increasing function of $s^{2}$.
	Accordingly,  $g_{-}(s)$ is a monotonically increasing function of $s$, because $g_{-}(s)>0$ and we assume that $s>0$ ($0<\theta<2\pi$).
	$g_{-}(s)$ takes its minimum value $g_{-}(0)=0$ at $s=0$ and its maximum value $g_{-}(1)=\sqrt{3}/2$ at $s=1$.
	Similarly, we find that $g_{+}(s)$ takes $g_{+}(0)=1$ as its maximum value and $g_{+}(1)=\sqrt{3}/2$ as the minimum value.
	Thus, in the case of $-1<c_{1}<-\sqrt{3}/2$, we found the possible range of $c_{1}$ as
	\begin{equation}
		-g_{+}(s)\leq c_{1}\leq -\sqrt{3}/2.
		\label{eq:ineq1}
	\end{equation}
	
	\subsection{case of $ -\frac{\sqrt{3}}{2} < c_{1} \leq 0$}
	\label{subsec2}
	
	When $ -\frac{\sqrt{3}}{2} < c_{1} \leq 0$, the inequality (\ref{eq:ineq0}) is trivially satisfied.
	Thus, this range is also a solution of the inequality.
	
	\subsection{case of $ 0 < c_{1} < \frac{\sqrt{3}}{2}$}
	\label{subsec3}
	
	When $0 < c_{1} < \frac{\sqrt{3}}{2}$, both sides of Eq. (\ref{eq:ineq0}) are negative.
	By squaring both sides, we obtain the following form of inequality:
	\begin{equation}
		16 c^{4}_{1} -8(2+s^{2})c^{2}_{1}+9 s^{2} \geq 0.
		\label{eq:ineq03}
	\end{equation}
	Note that the direction of the inequality is opposite to that in Eq. (\ref{eq:ineq01}).
	The solution of Eq. (\ref{eq:ineq03}) with respect to $c_{1}$ is given by
	\begin{equation}
		c_{1}\geq g_{+}(s),~~c_{1}\leq -g_{+}(s),~{\rm or}~-g_{-}(s)\leq c_{1}\leq g_{-}(s).
	\end{equation}
	Considering the conditions, $0<g_{-}(s)<\sqrt{3}/2$, $\sqrt{3}/2<g_{+}(s)<1$, and $0 < c_{1} <\sqrt{3}/2$, we find that
	\begin{equation}
		0<c_{1}<g_{-}(s).
	\end{equation}
	
	Summarizing the above three results, we obtain
	\begin{equation}
		-g_{+}(s)\leq c_{1}\leq g_{-}(s)
	\end{equation}
	as the solution of Eq. (\ref{eq:ineq0}) when $c_{n}\geq0$.
	Because $s^{2}=1-c^{2}$,
	the functions $g_{\pm}(s)$ is thus rewritten as a function of $c$ as follows
	\begin{equation}
		g_{\pm}(c)=\frac{\sqrt{2+s^{2}\pm\sqrt{s^{4}-5s^{2}+4}}}{2}=\frac{\sqrt{3-c^{2}\pm |c|\sqrt{3+c^{2}}}}{2}=\frac{\sqrt{3-c^{2}\pm c_{n}\sqrt{3+c^{2}}}}{2},
	\end{equation}
	where we use $|c|=|c_{n}|=c_{n}$.
	We now obtain the range of $c_{1}$,
	\begin{equation}
		-\frac{\sqrt{3-c^{2}+(-1)^{n}c\sqrt{3+c^{2}}}}{2}\leq c_{1} \leq \frac{\sqrt{3-c^{2}-(-1)^{n}c\sqrt{3+c^{2}}}}{2},
	\end{equation}
	which we have already shown as Eq. (\ref{eq:bound}).
	Similarly, we obtain the same range of $c_{1}$ as above for the case of $c_{n}<0$.
	Thus, we now have the possible range of $c_{1}$, regardless of the sign of $c_{n}$.

	\section{``Twin" composite pulses}
	\label{appendix4}
	
		\begin{figure}[h]
		\begin{center}
			\includegraphics[width=180mm]{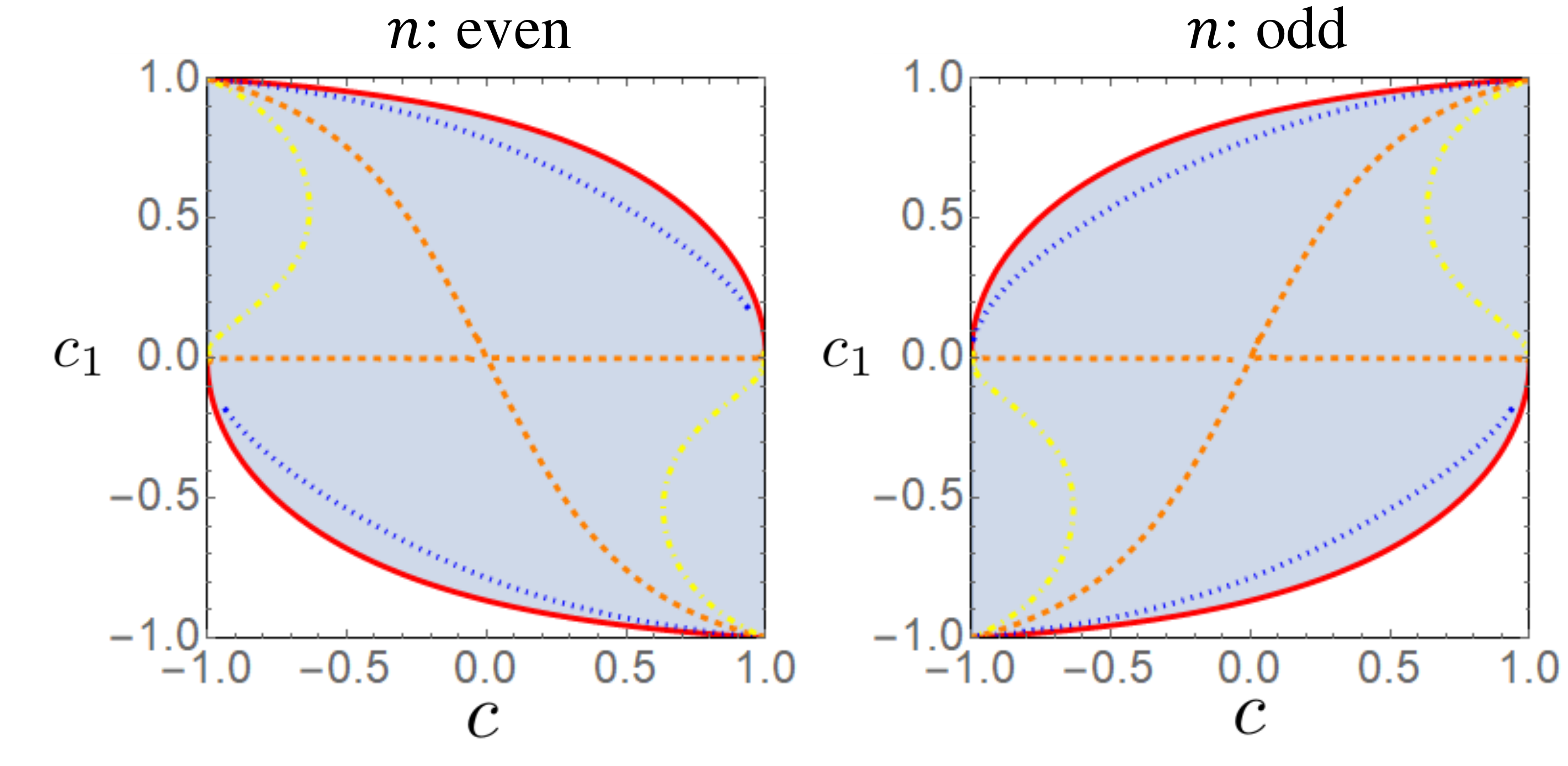}
			\caption{Contours with fixed $\alpha$. The solid, dotted, dashed, and dot-dashed lines represent $\alpha=-1.0,-0.3,0.5,$, and $0.7$, respectively.
				\label{fig:twin}}
		\end{center}
	\end{figure}	
	
	As mentioned in the main text, a fixed value of $\alpha$ provides two choices of $c_{1}$.
	We call these choices ``twins".
	The structure of twins is discussed here.
	Fig. \ref{fig:twin} shows the pairs of two lines which represent twins.
	The structure of these twins changes at $\alpha=0.5$.
	The twins are separated into the bottom and top parts when $\alpha<0.5$ and into the left and right parts when $\alpha>0.5$.
	In the case of $\alpha=0.5$, the twins touch each other at $c=0$.
		
	Paticularly, we have the explicit form of the twin of the CORPSE family:
	\begin{align}
		\theta^{(\rm tc)}_{1}&=2 \mu_{1}\pi-\theta/2+\kappa,\nonumber\\
		\theta^{(\rm tc)}_{2}&=2 \mu_{2}\pi+2 \kappa,\nonumber\\
		\theta^{(\rm tc)}_{3}&=2 \mu_{3}\pi-\theta/2+\kappa,\nonumber\\
		\phi^{(\rm tc)}_{1}&=\phi^{(\rm tc)}_{3}=\phi+\pi,\nonumber\\
		\phi^{(\rm tc)}_{2}&=\phi,
	\end{align}
	where $\kappa=\arcsin[\sin(\theta/2)/2]$ and $U(\theta,\phi)$ is the target operation.
	The integer parameters $\mu_{i}$'s satisfy $\mu_{1,3}\geq1$ and $\mu_{2}\geq0$.
	In our construction, the above operation sequence corresponds to the choice of $c_{1}=c_{1,n,-}$ for arbitrary $n_{i}$'s.
	Note that our parameters $n_{i}$'s can take $n_{i}=0$, unlike the case of the CORPSE family and its twin, owing to the choice of effective variables $\theta^{(p)}_{i}$'s.

	\section{Properties of gate infidelity}
	\label{appendix5}
	
	\subsection{order behaviour of gate infidelity}
	
	We discuss the order behaviour of the gate infidelity.
	One might expect that the gate infidelity $F$ would behave as $F\approx {\cal O}(f^{2})$ for (first-order) ORE-robust CPs because they are constructed such that the first-order error term in the unitary operator disappears.
	Actually, this is not the case.
	The gate infidelity for any quantum gate with ORE has the dependence of $F \approx {\cal O}(f^{2})$, regardless of its ORE robustness, whereas that for ORE-robust CPs has the dependence of $F\approx {\cal O}(f^{4})$.
	
	We can prove this fact not only for ORE but also for any type of error.
	To do this, it is convenient to introduce the Hamiltonian description of the error robustness. 
	The time evolution with a time-dependent Hamiltonian $H(t)$ is described by the Schr\"odinger equation,
	\begin{equation}
		\frac{d}{d t}|\Psi(t)\rangle= - i H(t) |\Psi(t)\rangle.
		\label{eq:schreq}
	\end{equation}
	The formal solution from $t=0$ to $t=T$ of this equation is
	\begin{equation}
		|\Psi(T)\rangle = \vt\exp \Bigl(- i \int^{T}_{0} d t H(t)\Bigr) |\Psi(0)\rangle,
		\label{eq:formalschr}
	\end{equation}
	where we introduce the time-ordered product:
	\begin{equation}
		\vt(A(t_{1})B(t_{2}))=
		\begin{cases}
			A(t_{1})B(t_{2})~~~~t_{1}>t_{2},\\
			B(t_{2})A(t_{1})~~~~t_{2}>t_{1}.
		\end{cases}
	\end{equation}
	The extension to cases of multiple operators, such as $\vt(A(t_{1})B(t_{2})C(t_{3}))$ is clear.
	When we consider a piecewise constant Hamiltonian, the time-ordered exponential in Eq. (\ref{eq:formalschr}) can be written as
	\begin{equation}
		\vt \exp \Bigl(-i \int^{T}_{0} d t H(t)\Bigr)=\exp \bigl(-i (t_{k}-t_{k-1}) H_{k}\bigr)\exp \bigl(-i (t_{k-1}-t_{k-2}) H_{k-1}\bigr)\cdots \exp \bigl(-i (t_{1}-t_{0}) H_{1}\bigr),
		\label{eq:piecewise}
	\end{equation}
	where $t_{0}=0$ and $t_{k}=T$, and the piecewise constant Hamiltonian is
	\begin{equation}
		H(t)=H_{i},~t_{i-1} \leq t < t_{i},~(i=1 \sim k).
		\label{eq:piecewise}
	\end{equation}
	Thus, we reproduce a sequence of elementary operations, as discussed in the main text.

	We consider the case where the Hamiltonian is decomposed into the following two parts:
	\begin{equation}
		H(t)=H_{0}(t)+ H_{\rm err}(t).
		\label{eq:hamiltonianwitherror}
	\end{equation}
	$H_{0}(t)$ represents the ideal Hamiltonian, whose dynamics $U_{0}(T,0):=\vt\exp(-i \int^{T}_{0}d t H_{0}(t))$ corresponds to the target operation $U(\theta,\phi)$ in the main text.
	We control this part $H_{0}(t)$ in the Hamiltonian.
	Meanwhile, $H_{\rm err}(t)$ is the effect of undesirable systematic errors during the operation. 
	The magnitude of $H_{\rm err}(t)$ is assumed to be sufficiently small during the operation.
	Either part $H_{0}(t)$ or $H_{\rm err}(t)$ can have time dependence.
	The state in the interaction picture with respect to $H_{0}(t)$ is given as
	\begin{equation}
		|\Psi_{I}(t)\rangle := \Bigl( \vt \exp \Bigl(- i \int^{t}_{0} d t' H_{0}(t')\Bigr) \Bigr)^{\dagger}|\Psi (t) \rangle=U^{\dagger}_{0}(t,0)|\Psi(t)\rangle,
		\label{eq:interactiondefinition}
	\end{equation}
	where $|\Psi(t)\rangle$ is a solution of the Schr\"odinger equation (\ref{eq:schreq}).
	The dynamics of the state $|\Psi_{I}(t)\rangle$ obeys the following equation:
	\begin{equation}
		\frac{d}{d t}|\Psi_{I}(t)\rangle=- i \tilde{H}_{\rm err}(t)|\Psi_{I}(t)\rangle,
		\label{eq:interaction}
	\end{equation}
	where $\tilde{H}_{\rm err}(t):=U^{\dagger}_{0}(t,0)H_{\rm err}(t)U_{0}(t,0)$.
	Its formal solution is
	\begin{equation}
		|\Psi_{I}(T)\rangle=\tilde{U}_{\rm err}(T,0)|\Psi_{I}(0)\rangle:= \vt\exp \Bigl(- i \int^{T}_{0} d t \tilde{H}_{\rm err}(t)\Bigr) |\Psi_{I}(0)\rangle.
		\label{eq:formalint}
	\end{equation}
	By comparing Eqs. (\ref{eq:formalint}) and (\ref{eq:interactiondefinition}), we obtain
	\begin{equation}
		|\Psi(T)\rangle=U_{0}(T,0)|\Psi_{I}(T)\rangle=U_{0}(T,0)\tilde{U}_{\rm err}(T,0)|\Psi(0)\rangle.
		\label{eq:importantrep}
	\end{equation}
	Thus, we formally decouple the effect of the error $\tilde{U}_{\rm err}(T,0)$ from the ideal dynamics $U_{0}(T,0)$.
	
	We intend to implement the ideal operation $U_{0}(T,0)$ accurately under the error $H_{\rm err}$.
	As the magnitude of $H_{\rm err}$ is sufficiently small, we ignore the higher-order terms in $\tilde{U}_{\rm err}(T,0)$ and obtain
	\begin{equation}
		\tilde{U}_{\rm err}(T,0) \sim \Bigl(1- i \int^{T}_{0} d t \tilde{H}_{\rm err}(t)\Bigr).
	\end{equation}
	The condition
	\begin{equation}
		\int^{T}_{0} d t \tilde{H}_{\rm err}(t)=0.
		\label{eq:errorobustcondition}
	\end{equation}
	implies that the operation (\ref{eq:importantrep}) is equal to the ideal $U_{0}(T,0)$ up to the first order of the magnitude of $\tilde{H}_{\rm err}(t)$.
	Assuming that both $H_{0}(t)$ and $H_{\rm err}(t)$ are piecewise constant,
	we reproduce the common description of error-robust CPs.
	
	Now, we show that the second-order term of an error in the gate infidelity disappears for any CP robust against the error.
	We first note that in one-qubit control, the first-order term of the gate infidelity disappears even for elementary operations.
	The gate infidelity is written as 
	\begin{equation}
		F:=1-\big|{\rm tr}\bigl( U^{\dagger}U_{0}(T,0)\tilde{U}_{\rm err}(T,0)\bigr) \big|/2=1-\big|{\rm tr}\bigl(\tilde{U}_{\rm err}(T,0)\bigr) \big|/2,
	\end{equation}
	where we use $U=U_{0}(T,0)$.
	The expansion of $F$ up to the first order gives
	\begin{equation}
		F\sim  \bigg|\int^{T}_{0}d t~{\rm tr} (\tilde{H}_{\rm err}(t))\bigg|+{\cal O}(f^{2}). 
	\end{equation}
	However, the trace of the Hamiltonian can always be taken as zero, and thus the first-order term of the gate infidelity always disappears regardless of the Hamiltonian.
	
	The second-order term can indicate whether the sequence is a CP or not.
	The second-order term of $F$ is given as
	\begin{equation}
		{\rm tr } \Biggl(\int^{T}_{0}d t \int^{t}_{0}d t'  \tilde{H}_{\rm err}(t)\tilde{H}_{\rm err}(t') \Biggr).
	\end{equation}
	This term can be rewritten as
	\begin{align}
		{\rm tr } \Biggl(\int^{T}_{0}d t \int^{t}_{0}d t'   \tilde{H}_{\rm err}(t)\tilde{H}_{\rm err}(t') \Biggr)=&{\rm tr}\Biggl( \frac{1}{2}\int^{T}_{0}d t \int^{t}_{0}d t' \tilde{H}_{\rm err}(t)\tilde{H}_{\rm err}(t') +\frac{1}{2}\int^{T}_{0}d t \int^{t}_{0}d t'  \tilde{H}_{\rm err}(t)\tilde{H}_{\rm err}(t')\nonumber\\
		&+\frac{1}{2}\int^{T}_{0}d t \int^{T}_{t}d t' \tilde{H}_{\rm err}(t)\tilde{H}_{\rm err}(t')-\frac{1}{2}\int^{T}_{0}d t \int^{T}_{t}d t' \tilde{H}_{\rm err}(t)\tilde{H}_{\rm err}(t')\Biggr)\nonumber\\
		=&	{\rm tr } \Biggl(\frac{1}{2} \int^{T}_{0}d t \int^{T}_{0}d t' \tilde{H}_{\rm err}(t)\tilde{H}_{\rm err}(t') + \frac{1}{2}\int^{T}_{0}d t \int^{t}_{0}d t'[\tilde{H}_{\rm err}(t),\tilde{H}_{\rm err}(t')]\Biggr)\nonumber\\
		=&	{\rm tr }\Biggl(\frac{1}{2}\Bigl(\int^{T}_{0}dt \tilde{H}_{\rm err}(t)\Bigr)^{2} \Biggr)+{\rm tr}\Biggl(\frac{1}{2}\int^{T}_{0}d t \int^{T}_{t}d t'[\tilde{H}_{\rm err}(t),\tilde{H}_{\rm err}(t')]\Biggr),
	\end{align}
	where we use $\int^{T}_{0}d t \int^{T}_{t}d t'=\int^{T}_{0}d t ' \int^{t'}_{0}d t$.
	The term $[\tilde{H}_{\rm err}(t),\tilde{H}_{\rm err}(t')]$ can be written as a linear combination of the Pauli matrices for any $t$ and $t'$, and is then traceless.
	Thus, the second term in the above equation disappears.
	The condition that the first term equals zero is equivalent to the first-order robustness of CPs, Eq. (\ref{eq:errorobustcondition}).
	Therefore, for any CP satisfying the robustness condition, the second-order term of the gate infidelity disappears.
	Thus, we can use the gate infidelity as a measure to characterize the performance of CPs by focusing on the second-order term.

	\subsection{gate infidelity as an upper bound of state infidelity}
	
	Here, we show that gate infidelity $1-|{\rm tr}(U^{\dagger}V)|/2$ is an upper bound of the state infidelity $1-|\langle \psi |U^{\dagger}V|\psi\rangle|$ for any state $|\psi \rangle$ and any $2\times 2$ $SU(2)$ matrices $U$ and $V$.
	Note that the dynamics by the traceless Hamiltonian, as in the main text, is always expressed by an $SU(2)$ matrix.  
	Instead of treating the gate and state infidelities directly, we simply compare the gate fidelity $|{\rm tr}(U^{\dagger}V)|/2$ with the state fidelity $|\langle \psi |U^{\dagger}V|\psi\rangle|$ for any state $|\psi \rangle$.
	Because $U^{\dagger}V$ is also a $SU(2)$ matrix, we can diagonalize this.
	The eigenvalues are denoted by $e^{i g}$ and $e^{-i g}$, whereas the corresponding eigenvectors are $|\psi_{+}\rangle$ and $|\psi_{-}\rangle$.
	The square of the gate fidelity is then calculated as
	\begin{equation}
		|{\rm tr}(U^{\dagger}V)|^{2}/4=|e^{i g}+e^{-i g}|^{2}/4=\cos^{2} g.
		\label{eq:cos}
	\end{equation}
	On the other hand, as any state $|\psi \rangle$ can be written as $|\psi\rangle=a |\psi_{+}\rangle+b|\psi_{-}\rangle$ $(|a|^{2}+|b|^{2}=1)$, we can calculate the square of the fidelity for a state as
	\begin{align}
		|\langle \psi | U^{\dagger}V | \psi \rangle|^{2}=&\Big||a|^{2}\langle \psi_{+} | U^{\dagger}V | \psi_{+} \rangle+|b|^{2}\langle \psi_{-} | U^{\dagger}V | \psi_{-}\rangle+a^{*}b\langle \psi_{+} | U^{\dagger}V | \psi_{-}\rangle+a b^{*}\langle \psi_{-} | U^{\dagger}V | \psi_{+} \rangle  \Big|^{2}\nonumber\\
		=&\Big||a|^{2}e^{i g}+|b|^{2}e^{-i g}  \Big|^{2}=1-2|a|^{2}|b|^{2}(1-\cos(2 g))=1-4|a|^{2}(1-|a|^{2})\sin^{2} g.
		\label{eq:sin}
	\end{align} 
	By subtracting the squared gate fidelity from the squared fidelity, we obtain
	\begin{align}
		|\langle \psi | U^{\dagger}V | \psi \rangle|^{2}-|{\rm tr}(U^{\dagger}V)|^{2}/4=&\sin^{2} g-4|a|^{2}(1-|a|^{2})\sin^{2} g \nonumber\\
		=&(1-4|a|^{2}(1-|a|^{2}))\sin^{2}g.
	\end{align}
	We can easily verify that the term in the parentheses is always positive because the maximum of $|a|^{2}(1-|a|^{2})$ is $1/4$.
	Thus, we have shown that the state fidelity for any state is bounded below by the gate fidelity.
	Accordingly, the state infidelity for any state is bounded from above by the gate infidelity.

	\section{Performance evaluation via state infidelity}
	\label{appendix6}
	
	As mentioned in the main text, the CORPSE family does not necessarily have the best accuracy when we consider measures other than the gate infidelity.
	For example, we evaluate the state infidelity $F_{\rm state}:=1-\big|\langle\psi |U^{\dagger}U'|\psi \rangle\big|$, where $U'$ is an operation with ORE and $U$ is the corresponding errorless operation.
	$|\psi\rangle$ is the initial state of the operation; hence, $F_{\rm state}$ represents the difference between the final states with and without ORE.
	We consider $|\psi\rangle=|0\rangle$, which is an eigenstate of $\sigma_{z}$ with eigenvalue $1$: $\sigma_{z}|0\rangle=|0\rangle$.
	
\begin{figure}[h]
	\begin{center}
		\includegraphics[width=180mm]{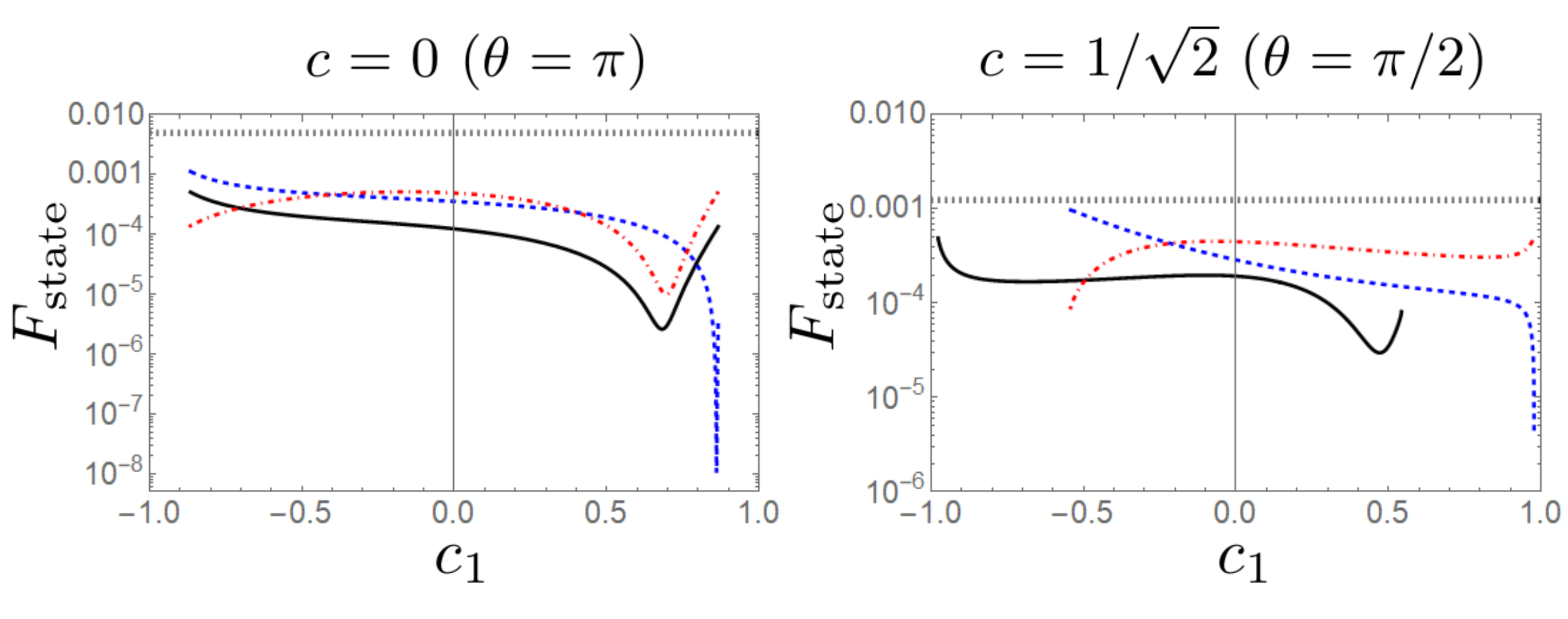}
		\caption{State infidelity for the cases of $c=1/\sqrt{2}~(\theta=\pi/2)$ and $c=0 ~(\theta=\pi)$.
			The initial state is taken to be $|0 \rangle$.
			The solid, dashed, and dot-dashed curves correspond to $(n_{1},n_{2},n_{3})=(0,0,0)$, $(n_{1},n_{2},n_{3})=(1,0,0)$, and $(n_{1},n_{2},n_{3})=(0,1,0)$, respectively.
			The dotted straight line represents the state infidelity of the elementary operation with ORE.
			The error parameter is set to $f=0.1$.
			All curves for $c=0$ have the same existence range because $c(1,n,\pm)=\pm \sqrt{3}/2$, regardless of $n$ in this case.
			On the other hand, the curve corresponding to $(n_{1},n_{2},n_{3})=(0,0,0)$ for the case of $c=1/\sqrt{2}$ has a different existence range than the other curves: $c_{1,n,\pm}$ depend on $n$ unless $c=0$.
			\label{fig:state_fidelity}}
	\end{center}
	\end{figure}

	Fig. \ref{fig:state_fidelity} shows the state infidelity with several combinations of $(n_{1},n_{2},n_{3})$ for the cases of $c=1/\sqrt{2}~(\theta=\pi/2)$ and $c=0 ~(\theta=\pi)$ in the same way as Fig. \ref{fig:fidelity}.
	Evidently, the state infidelity for several cases of $(n_{1},n_{2},n_{3})$ takes its minimal value at a point that is neither the left nor right end.
	Thus, the CORPSE family and its twins do not necessarily have the best accuracy for fixed $(n_{1},n_{2},n_{3})$ when we take the state fidelity as an accuracy measure.
	The minimum state infidelity among the evaluated combinations of $(n_{1},n_{2},n_{3})$ appears on the curve representing $(n_{1},n_{2},n_{3})=(1,0,0)$ as in the case of the gate fidelity (Fig. \ref{fig:fidelity}) although the corresponding sequence is not fundamental CORPSE when $c=0~(\theta=\pi)$.
	
\end{document}